\documentclass[aps,pra,showpacs,twoside,twocolumn,10pt]{revtex4-1}
\usepackage[colorlinks=true, citecolor=red, urlcolor=blue ]{hyperref}
\usepackage{times,epsfig,amssymb,amsfonts,amsmath,bm,subfigure,mathtools,amsthm,braket,soul,enumitem,color,physics,graphics,graphicx}
\usepackage[normalem]{ulem}

\begin{document}

\title{Circulating  Genuine Multiparty Entanglement in Quantum Network
}
\author{Pritam Halder$^{1}$, Ratul Banerjee$^{1}$, Srijon Ghosh$^{1}$, Amit Kumar Pal$^{2}$, Aditi Sen (De)$^{1}$}
\affiliation{$^1$Harish-Chandra Research Institute, A CI of Homi Bhabha National Institute, Chhatnag Road, Jhunsi, Prayagraj 211019, India}
\affiliation{$^2$ Department of Physics, Indian Institute of Technology Palakkad, Palakkad 678 623, India}

\begin{abstract}

We propose a deterministic scheme of generating genuine multiparty entangled states in quantum networks of arbitrary size having various geometric structures -- we refer to it as \emph{entanglement circulation}. The procedure involves optimization over a set of two-qubit arbitrary unitary operators and the  entanglement of the initial resource state. We report that the set of unitary operators that maximize the genuine multipartite entanglement quantified via generalized geometric measure (GGM) is not unique. We prove that the GGM of the resulting state of arbitrary qubits coincides with the minimum GGM of the initial resource states. By fixing the output state as the six-qubit one, we find the optimal way to create such states according to the available resource.  
Moreover, we show that the method proposed here can be implemented by using logic gates, or by using the time dynamics of  realizable spin Hamiltonians.  In case of an ordered system, GGM varies periodically with time while  the evolution via disordered models lead to a low but constant multipartite entanglement in outputs at a critical time, which decreases exponentially with the increase of the strength of the disorder.  
\end{abstract}

\maketitle

\section{Introduction} 

Distributing information with minimal errors between several parties (nodes) situated in distant locations remains a challenging problem both in the classical and quantum domains~\cite{Kimble2008}. In recent times, quantum networks with promising applications in fields ranging from secure communication~\cite{Poppe2008}, exponential gains in communication complexity~\cite{Guerin2016}, and clock synchronization~\cite{Komar2014} to distributed quantum computing~\cite{cirac1999} and distributed function computing~\cite{Rodney2011,Claude2002} have become an active direction of research. It has also been established that similar to the initial proposals of quantum communication schemes  with a single sender and a single receiver ~\cite{ekert1991, bennett, bennett1992, bennett1993} requiring bipartite entanglement as resource~\cite{Horodecki2009}, multipartite entanglement  can be a key ingredient in most of the quantum information processing tasks in quantum networks. Therefore, creating/generating a multipartite state via appropriate quantum operations from states having lesser number of parties  with the assurance of multipartite entanglement is one of the important enterprises in the development of quantum networking test-beds. 


Towards creating a quantum network having genuine multipartite entanglement,  several proposals have been developed in the last few years.  
There are broadly two methods by which genuine multiparty entangled  states can be shared over a quantum network. One of them is the probabilistic creation of genuine multiparty entangled states \cite{Acin2007,Rausendorf2001,Walther2005,Briegel2009},  using  either projective, or unsharp,  or positive operator valued measurements~\cite{Krauss1983}, while the other one is the deterministic process  engaging unitary operations, or logic gate implementations  in a quantum circuit~\cite{nielsen_chuang_2010}. Specifically, starting from several copies of noisy states,  one can setup a quantum network by employing  quantum repeaters~\cite{briegel1998,dur1999}, which is either a combination of entanglement distillation~\cite{bennett1996} and swapping~\cite{zukowski1993,bose1998}. It was later generalized to the measurement-based scheme on lattices of  different  dimensions \cite{Wallnofer2016}. A multiparty entangled state can also be grown by performing projective measurements in a star network~\cite{sen(De)2005,Cavalcanti2011,Banerjee2020}, or by applying unsharp measurement on a single party of the multipartite state and an auxiliary qubit ~\cite{Halder2021}. On the other hand, there have also been propositions and experimental demonstrations of several techniques such as fusion and expansion producing large  multipartite entangled states eg. Greenberger Horne Zeilinger (GHZ)~\cite{Greenberger2007}, \(W\)~\cite{Durvidal2002}, and cluster states~\cite{Briegel2001}, starting from small  entangled states~\cite{zang2015, sharma2020, Severin2021,WaltherZeilinger2005,ZeilingerHorne1997,Browne2005,Tashima(1)2009,Tashima(2)2009,Tashima(3)2008,Tashima(4)2011,Ozaydin2014,Ozdemir2011,Bugu2013}. However, notice that in most of these  works, the network-building mechanism have been constructed to create a specific class of multipartite states which are known to be important for quantum computation or quantum communication tasks~\cite{Kozlowski2019,Azuma2021,Pirker2018,Pirker2019,Meignant2019,Gyongyosi2019,Miguel2020,Miguel2021}.  

Going beyond the  realm of creating specific entangled states,  in this work, we provide a generic method to \emph{deterministically} generate multipartite entangled states. In particular, we address the following questions: 
\begin{enumerate}
\item \emph{Does a protocol for designing a quantum network with a fixed genuinely multipartite entanglement content, having a fixed size and a  geometry, exist?}

\item \emph{If such a protocol exists, is there an optimal resource?}

\item \emph{Is this protocol robust against imperfections in the required operations}?
\end{enumerate}
In this work, we answer all three of these questions affirmatively. We provide a protocol for distributing genuinely multiparty entangled states with a fixed generalized geometric measure (GGM) \cite{aditi2010} over a large quantum network of fixed number of parties and of particular geometry -- we call it as \emph{entanglement circulation}. Moreover, we identify optimal resource states according to the amount of multiparty entanglement present in the output states, and optimal two-qubit unitary operators that can be implemented  in terms of single and two qubit logic gates  in circuit models~\cite{Barenco1995, Farrokh2004,Williams2004}. We also present a variant of this protocol where the output entangled state can be generated by a time-evolution governed by a chosen quantum many-body Hamiltonian with or without disorder, which is realizable by currently available technologies based on photons~\cite{photonfrus}, and trapped ions~\cite{abah2012, low2020, puebla2019, erhard2018}.

\begin{figure}
\includegraphics[width=\linewidth]{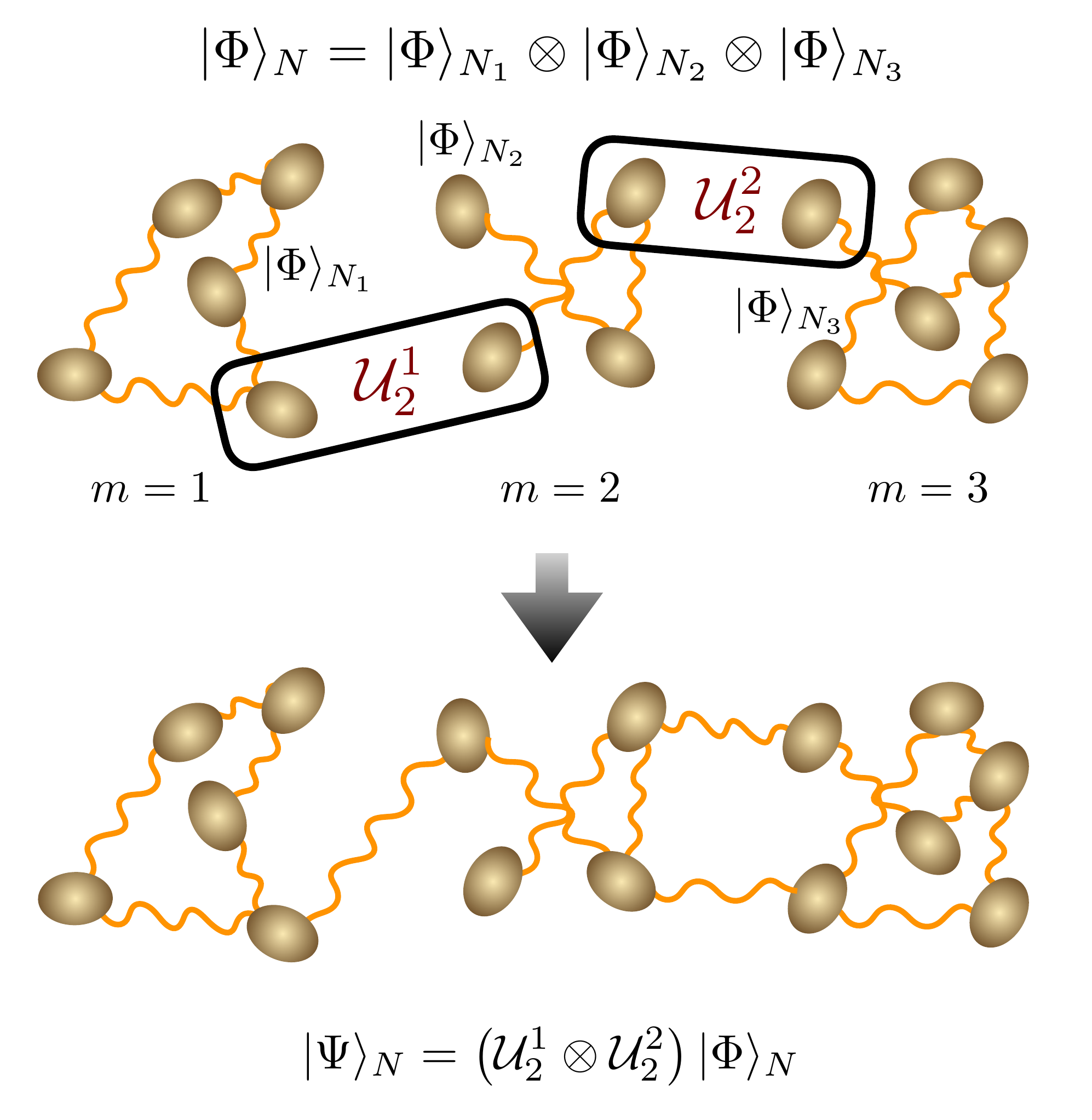}
\caption{(Color online.) Schematic representation of the \emph{entanglement circulation protocol} (ECP) for constructing large multi-qubit states by applying arbitrary two-qubit unitary operation over a number of smaller unit states (see Sec.~\ref{sec:general_protocol}). Here, three unit cells of qubits, constituted of $N_1$, $N_2$, and $N_3$ qubits, are combined by a unitary operator $\mathcal{U}_2^1\otimes\mathcal{U}_2^2$ with $M=4$ to create a larger multi-qubit system of $N=N_1+N_2+N_3$ qubits. Our aim is to find out the optimal unitary operator,  $\mathcal{U}_2^1\otimes\mathcal{U}_2^2$  such that the resulting multipartite state possess maximal genuine multipartite entanglement. }
\label{fig:schem1}
\end{figure}

Precisely, by considering $m$ number of initially entangled states, each comprised with $N_i$, $i=1,2,\cdots,m$ number of qubits, we apply  $L (\leq m)$  number of optimized two-qubit unitary operators acting on one qubit of each of the $N_i$-party state (see Fig. \ref{fig:schem1}), so that a genuine multiparty entangled state having $N=\sum_{i=1}^mN_i$ parties is produced. Existence of genuine multiparty entanglement in the newly created state is confirmed by computing the generalized geometric measure ~\cite{wei2003, wei2005_a, blasone2008, orus2008, aditi2010}. We prove that to establish a network consisting of $N$ parties with a fixed amount of GGM, one needs to create a resource state having lesser number of parties containing at least the same amount of GGM, since the minimum entanglement among the resource states  coincides with the GGM of the resulting state,  obtained after applying optimal unitary operators on the resource states. Starting with  arbitrary three-qubit pure states as the resource, we provide a recursion relation for obtaining the network with a large number of parties. In this scenario, we also identify the optimal region in the parameter space of the unitary operators to obtain the maximum possible entanglement in the resultant state. To assess the effectiveness of the method, we generate  multipartite initial states in a Haar uniform  way  and demonstrate that there is a trade-off  between the initial resource states of $N_1$ and $N_2$-parties, and the resulting state having $N_1+N_2$ parties. 

In situations where the two-qubit unitary operation is dictated by an interacting ordered spin Hamiltonian instead of an arbitrary unitary operators, our analysis again presents  the recursion relation for the final state with $3m$ parties starting from  $m$ copies of three-qubit initial entangled resource states.  We also determine the optimal time in which the maximal GGM can be created from a given initial state by investigating the dynamics of the multipartite entanglement of the resulting state. Extending our investigation into non-ideal scenarios where disorder can naturally appear in the interaction strength of the spin Hamiltonian~\cite{Lewenstein2007, Aspect08, deMartini2008, Aspect2009},  we show that even in the presence of disorder in the operation, few copies of initially entangled states with lesser number of parties can lead to a finite amount of quench-averaged genuine multipartite entanglement in the output state. Interestingly, we observe that  in the ordered case,  maximal entanglement can only be produced during certain time intervals, while for evolution governed by a disordered spin models,  quench-averaged genuine multipartite entanglement in the output state remains almost constant after a  certain critical time. Furthermore, we provide a prescription for obtaining  multipartite entangled states over a quantum network that is  obeying a triangular geometry.

The rest of our paper is presented as follows. In Sec. \ref{sec:general_protocol},  we introduce the procedure to extend multiparty entangled states over large quantum networks and prove  bounds on entanglement of the output state in terms of the entanglement in the resource states. We present a recursion relation in Sec. \ref{sec:three_qubit_units} where thee-qubit states are used as inputs, and investigate the features of the two-qubit unitary operations as well as the optimal distribution of the resource states over the quantum network for obtaining the desired output state. We also comment on the possibility of growing the output quantum state following a triangular geometry in SubSec. \ref{subsec:geometry} . In Sec.~\ref{sec:many_body}, we explore the possibility of obtaining the desired genuinely multiparty entangled state over a quantum network of fixed number of parties as a result of a time evolution governed by a quantum many-body Hamiltonian, and discuss the effect of the presence of disorder in the Hamiltonian on the output state. The concluding remarks can be found in Sec.~\ref{conclusion}.

\section{Entanglement Circulation protocol}
\label{sec:general_protocol}

We now introduce a procedure for preparing large multiparty entangled states, starting from a number of genuine multiparty entangled states of small number of qubits, using unitary operations. Let us consider a system of $N=\sum_{i=1}^m N_i$ qubits constituted of $m$ disconnected groups of qubits, where the  group $i$ has $N_{i}$ qubits. Each group of qubits is represented by an entangled state $\ket{\Phi}_{N_i}$, such that the initial state $\ket{\Phi}_N$ of the $N$-qubit system is given by   
\begin{equation}
   |\Phi\rangle_N = \bigotimes_{i = 1}^m |\Phi \rangle_{N_{i}}. 
   \label{eq:initial state}
\end{equation}
An $M$-qubit unitary operator $\mathcal{U}_M$ is operated on the state $\ket{\Phi}_N$, where the $M$-qubit support of $\mathcal{U}_M$ is constituted of taking at least one qubit from each of the $m$ groups of qubits, i.e., $m\leq M\leq N$ (see Fig.~\ref{fig:schem1}).  For the ease of discussion, we refer to these groups of qubits to be the \emph{unit cells}, and  the corresponding states $\{\ket{\Phi}_{N_i}\}$ to be the \emph{unit states}.  
Note that  the unit states can be considered to have identical sizes in typical quantum network building exercises, so that \(N_i = N_j\), \((i\neq j)\), although  we will also deal with unit states of different sizes, i.e., \(N_i \neq N_j\), \((i\neq j)\),  as demonstrated in the following subsection.
The resultant $N$-qubit pure state reads as
\begin{eqnarray}
\ket{\Psi}_{N}=\mathcal{U}_M \otimes I_{\overline{M}} \ket{\Phi}_N,
\label{eq:final_state}
\end{eqnarray}
which depends on the parameters involved in the quantum states $\{\ket{\Phi}_{N_i},i=1,2,\cdots,m\}$ as well as $\mathcal{U}_M$, where $I_{\bar{M}}$ denotes the identity operator in the qubit Hilbert space acted on the rest of the qubits except \(M\), we denote the set by \(\bar{M}\). We claim that a unitary operation $\mathcal{U}_M$ and the suitable initial state can lead to a genuine multipartite entangled state, $\ket{\Psi}_N$, as will be shown in subsequent section. In a network, a node may be a collection of qubits made by taking at least one from each of a subset of unit cells and 
$\mathcal{U}_M$ can act on a subset of the set of qubits constituting a \emph{node} of the network. We refer this method of creating multipartite entangled state  as the \emph{entanglement circulation protocol} (ECP). Since any arbitrary unitary operators can be decomposed in terms of single and two qubit logic gates \cite{Barenco1995},  repetitive applications of these gates can implement \(\mathcal{U}_M\), thereby create multiparty entangled states \cite{Farrokh2004,Williams2004}. Notice that since the ECP is based on unitary operations, the number of parties in the process is conserved, which may not be the case for some of the measurement-based protocols~\cite{briegel1998, Wallnofer2016}.

In situations where any of the $m$ genuinely multiparty entangled states is $k$-separable (see Appendix~\ref{sec:ggmdef}), say, $\ket{\Phi}_{N_i}$,  this protocol can be applied to first create $N_i$-qubit genuine multiparty entangled state by designing an appropriate unitary operator $\mathcal{U}_{M^\prime}$ ($M^\prime\leq N_i$), and  subsequently apply \(\mathcal{U}_M\)  to create the $N$-qubit genuine multiparty entangled state following Eq.~(\ref{eq:final_state}). Note also that although we discuss the protocol in detail with arbitrary multi-qubit pure states in the subsequent sections, the protocol has the potential for generalization to mixed states of qubits as well as in higher-dimensional systems.

In this work, to quantify  genuine multipartite entanglement of the resulting state,   a distance-based  entanglement measure, namely the generalized geometric measure  (GGM)  (see Appendix~\ref{sec:ggmdef} for a definition)  is computed \cite{aditi2010}.   

\subsection{Bounding GGM of  output states with GGMs of inputs}
\label{subsec:protocol}

Let us illustrate the  multipartite entanglement circulation protocol described above with the minimal support for the unitary operator by fixing $M=2$. An arbitrary two-qubit unitary operator $\mathcal{U}_2$ can be written as
\begin{equation}
    \mathcal{U}_2 = (A_1\otimes A_2) U_d (A_3\otimes A_4),
    \label{general_unitary}
\end{equation}
with $\{A_{i} \in U(2),i=1,2,3,4\}$, and $U_{d}$ being a ``non-local" component of the  operator, given by
\begin{eqnarray}
U_{d} = \exp\left[-\text{i}\sum_{j=x,y,z}\alpha_{j}\sigma_{j} \otimes \sigma_{j}\right],
\label{unitary_d}
\end{eqnarray}
where $0 \leq \alpha_{j} \leq \frac{\pi}{2}$, $\sigma_{j}$ are the Pauli matrices, and $\alpha_{j} \in \mathbb{R}$, for $j = x, y, z$. Note that $A_1$ and $A_2$ being local unitary operators that keep entanglement unchanged, an implementation of $U_d$ is sufficient to carry out the proposed protocol, which can be achieved by a non-trivial combination of five elementary single-qubit (rotation) gates and three CNOT gates \cite{Farrokh2004}. These local rotations  introduce the entanglement controlling parameters $(\alpha_x, \alpha_y, \alpha_z)$ in the ECP.

Let us first consider two arbitrary unit states, $\ket{\Phi}_{N_1}$ and $\ket{\Phi}_{N_2}$, with $N_1,N_2\geq 2$,  such that the initial state $\ket{\Phi}_N$ having $N=N_1+N_2$ qubits can be represented as
\begin{eqnarray}
\ket{\Phi}_N=\ket{\Phi}_{N_1}\otimes\ket{\Phi}_{N_2}.
\end{eqnarray}
The resultant state, 
\begin{eqnarray}
\ket{\Psi}_N=\mathcal{U}_2\ket{\Phi}_N
\end{eqnarray} 
is obtained by applying the two-qubit unitary operator $\mathcal{U}_2$ on a qubit-pair constituted of one qubit from each of the unit cells. Let  the GGMs of the two units states $\ket{\Phi}_{N_1}$ and $\ket{\Phi}_{N_2}$ be $\mathcal{G}_1$ and $\mathcal{G}_2$ respectively, while the GGM of the final state $\ket{\Psi}_N$ is given by $\mathcal{G}$. We present Proposition I that conditionally expresses $\mathcal{G}$ in terms of $\mathcal{G}_1$ and $\mathcal{G}_2$. 

\noindent\textbf{$\blacksquare$ Proposition I.} \emph{The GGM of arbitrary pure multi-qubit state, $\ket{\Psi}_N$, resulting from two arbitrary pure unit states $\ket{\Phi}_{N_1}$ and $\ket{\Phi}_{N_2}$ via application of optimal unitary operator, $\mathcal{U}_2$ on two qubits, one from each of the unit states, turns out to be
\begin{eqnarray}
\max_{\{\mathcal{U}_2\}}\mathcal{G}=\min\{\mathcal{G}_1,\mathcal{G}_2\},
\end{eqnarray}
with the condition that the values of $\mathcal{G}$, $\mathcal{G}_1$, and $\mathcal{G}_2$ correspond to the eigenvalue of any one of the single-qubit reduced density matrices obtained respectively from $\ket{\Psi}_N$, $\ket{\Phi}_{N_1}$, and $\ket{\Phi}_{N_2}$, by tracing out the rest of the qubits from them.}

\noindent\textbf{Proof.} We assume here that the GGM  for any arbitrary multi-qubit quantum state always corresponds to one of the single-qubit reduced density matrices computed from the quantum state by tracing out the rest of the qubits except one. Hence  we write the Schmidt decomposition of an arbitrary $N_1$-qubit pure state considering the  bipartition between a single qubit and the rest  as
\begin{eqnarray}
    \ket{\Phi}_{N_1} = \sqrt{\gamma_{1}}\ket{x}\ket{0}+\sqrt{\delta_{1}}\ket{y}\ket{1},
    \label{eq:unit_state_ggm_1}
\end{eqnarray}   
where $\{\ket{0},\ket{1}\}$ is the computational basis for the two-dimensional  Hilbert space. Similarly, one can also write, for an $N_2$-qubit state, 
\begin{eqnarray}
     \ket{\Phi}_{N_2} = \sqrt{\gamma_{2}}\ket{0}\ket{u}+\sqrt{\delta_{2}}\ket{1}\ket{v},  
     \label{eq:unit_state_ggm_2}
\end{eqnarray}
and the joint initial state of $N=N_1+N_2$ qubits reads as
\begin{eqnarray}
     \ket{\Phi}_{N} &=&  \sqrt{\gamma_{1} \gamma_{2}} \ket{x00u} + \sqrt{\gamma_{1} \delta_{2}} \ket{x01v}\nonumber \\ &&+ \sqrt{\delta_{1} \gamma_{2}} \ket{y10u} + \sqrt{\delta_{1} \delta_{2}} \ket{y11v}.
\end{eqnarray}
The resultant state $\ket{\Psi}_N=\mathcal{U}_2\ket{\Phi}_N$ is obtained by applying the arbitrary two-qubit unitary operator $\mathcal{U}_2$ on a pair of qubits which is constituted of one qubit from each of $\ket{\Phi}_{N_1}$ and $\ket{\Phi}_{N_2}$. However, note that the local unitary operators $\{A_i,i=1,2,3,4\}$ have no effect on entanglement, and therefore can be ignored. The non-local unitary, $U_d$, when expanded, takes the form 
\begin{equation}
 U_{d} = 
\begin{pmatrix}
\mu_{1} & 0 & 0 & \mu_{2} \\
0 & \mu_{3} & \mu_{4} & 0\\
0 & \mu_{4} & \mu_{3} & 0\\
\mu_{2} & 0 & 0 & \mu_{1} \\
\end{pmatrix},   
\label{unitary_d_matrix}
\end{equation}
where 
\begin{eqnarray} 
\mu_{1}&=&\text{e}^{-\text{i} \alpha_{z}} \cos(\alpha_{x}-\alpha_{y}), 
\mu_{2}=-\text{i} \text{e}^{-\text{i} \alpha_{z}}\sin(\alpha_{x}-\alpha_{y}),\nonumber \\ 
\mu_{3}&=&\text{e}^{\text{i} \alpha_{z}}\cos(\alpha_{x}+\alpha_{y}), 
\mu_{4}=-\text{i} \text{e}^{\text{i} \alpha_{z}}\sin(\alpha_{x}+\alpha_{y}).
\end{eqnarray} 
Determining the explicit effects of $U_d$ on the two-qubit computational basis as  
\begin{eqnarray}
U_{d}\ket{00} &=& \mu_1\ket{00}+\mu_2\ket{11},\;
U_{d}\ket{01} = \mu_3\ket{01}+\mu_4\ket{10},\nonumber\\
U_{d}\ket{10} &=& \mu_3\ket{10}+\mu_4\ket{01},\;
U_{d}\ket{11} = \mu_1\ket{11}+\mu_2\ket{00},\nonumber \\
\end{eqnarray}
and applying $U_d$ on two parties having dimension $2$ in the state $\ket{\Phi}_N$, each from one of $\ket{\Phi}_{N_1}$ and $\ket{\Phi}_{N_2}$, the resultant state of the joint system can be written as 
\begin{eqnarray}
\ket{\Psi}_{N} &=& \sqrt{\gamma_{1} \gamma_{2}} (\mu_{1} \ket{x00u} + \mu_{2} \ket{x11u}) \nonumber\\
&&+ \sqrt{\gamma_{1} \delta_{2}} (\mu_{3} \ket{x01v} + \mu_{4} \ket{x10v}) \nonumber\\
&&+ \sqrt{\delta_{1} \gamma_{2}} (\mu_{3} \ket{y10u} +  \mu_{4} \ket{y01u}) \nonumber \\ && + \sqrt{\delta_{1} \delta_{2}} (\mu_{1} \ket{y11v} +  \mu_{2} \ket{y00v}).
\label{eq:effective_four_party state}
\end{eqnarray}
Let us assume, without any loss of generality, that $\gamma_{1} \geq \delta_{1}$, and $\gamma_1 \geq \gamma_{2} \geq  \delta_{2}$, which implies $\min\{\mathcal{G}_{1},\mathcal{G}_2\} = 1 - \gamma_{1}$. Note also that $\ket{\Psi}_N$ is written as an effective four-party state in Eq.~(\ref{eq:effective_four_party state}), implying that considering all possible $1:\text{rest}$ and $2:\text{rest}$ bipartitions is sufficient for computing the GGM of the state $\ket{\Psi}_N$, $1$ and $2$ denoting the number of parties. Let us denote the single- and two-party density matrices, corresponding respectively to the $1:\text{rest}$ and $2:\text{rest}$ bipartitions, by $\rho_1^i$ ($i=1,2,3,4$) and $\rho_{2}^{ij}$ ($i=1; j=2,3,4$), and the corresponding maximum eigenvalues by $\lambda_1^i$ and $\lambda_2^{ij}$. Maintaining the assumption that the GGM of the state $\ket{\Psi}_N$ comes from the single-party reduced density matrices, we investigate the density matrices $\rho_1^i$s only. Clearly, 
\begin{eqnarray}
\lambda_1^1 &=& \max[\gamma_1 , 1- \gamma_1],\nonumber \\
\lambda_1^2 &=& \max[\epsilon_1 , 1- \epsilon_1],\nonumber \\ 
\lambda_1^3 &=& \max[\epsilon_2 , 1- \epsilon_2],\nonumber \\
\lambda_1^4 &=& \max[\gamma_2 , 1- \gamma_2], 
\end{eqnarray}
with
\begin{eqnarray}
\epsilon_1 &=& \frac{1}{2}\big[1 + (\gamma_1 +\gamma_2 -1)\cos2(\alpha_x -\alpha_y)\nonumber \\ 
&&+ (\gamma_1 - \gamma_2)\cos2(\alpha_x + \alpha_y)\big], \nonumber \\ 
\epsilon_2 &=& \frac{1}{2}\big[1 - (\gamma_1 +\gamma_2 -1)\cos2(\alpha_x -\alpha_y) \nonumber \\
&&+ (\gamma_1 - \gamma_2)\cos2(\alpha_x + \alpha_y)\big].  
\end{eqnarray}
Let us first assume $\alpha_z$ to be constant, and focus on the behaviors of $\epsilon_{1(2)}$ on the $(\alpha_x,\alpha_y)$ plane. The determinant of the Hessian for $\epsilon_{1(2)}$ can be constructed as
\begin{eqnarray}
\det \mathcal{H} &=& \begin{vmatrix}
\frac{\partial^2 \epsilon_{1,2}}{\partial \alpha_x ^2} & \frac{\partial^2 \epsilon_{1,2}}{\partial \alpha_x \partial \alpha_y} \\  
 \frac{\partial^2 \epsilon_{1,2}}{\partial \alpha_y \partial \alpha_x}& \frac{\partial^2 \epsilon_{1,2}}{\partial \alpha_y ^2} 
\end{vmatrix}.
\end{eqnarray}
From here onward, we present calculations only for $\epsilon_{1}$, while the calculations for $\epsilon_{2}$ are similar. Upon computation of the derivatives for $\epsilon_{1}$, the determinant of the Hessian for $\epsilon_{1}$ reads 
\begin{eqnarray} 
\det \mathcal{H}  &=& 8(\gamma_1 - \gamma_2) (-1 + \gamma_1 + \gamma_2) [\cos 4 \alpha_x + \cos 4 \alpha_y].\nonumber \\
\end{eqnarray}
Note that under the conditions, $\gamma_{1} \geq \delta_{1}$ and $\gamma_1 \geq \gamma_{2} \geq  \delta_{2}$, $\gamma_{1}>\gamma_2>1/2$. Within the range $0\leq \alpha_{x},\alpha_y\leq \pi$, the local maximums of $\epsilon_{1}$ are denoted by 
\begin{eqnarray}
\frac{\partial \epsilon_1}{\partial \alpha_x}=\frac{\partial \epsilon_1}{\partial \alpha_y}=0, \frac{\partial^2 \epsilon_1}{\partial \alpha_x ^2}<0,\det \mathcal{H}>0,
\end{eqnarray}
which occurs at $(0,0)$, $(0,\pi)$, $(\pi,0)$, $(\frac{\pi}{2},\frac{\pi}{2})$, and $(\pi,\pi)$. At these points, $\epsilon_1=\gamma_1$. On the other hand, the saddle points of the $\epsilon_1$ landscape over the $(\alpha_{x},\alpha_y)$ plane are denoted by 
\begin{eqnarray}
\frac{\partial \epsilon_1}{\partial \alpha_x}=\frac{\partial \epsilon_1}{\partial \alpha_y}=0, \frac{\partial^2 \epsilon_1}{\partial \alpha_x ^2}<0, \det \mathcal{H}<0,
\end{eqnarray}
which occurs at the points $(\frac{\pi}{4},\frac{\pi}{4})$, $(\frac{\pi}{4},\frac{3\pi}{4})$, $(\frac{3\pi}{4},\frac{\pi}{4})$, and $(\frac{3\pi}{4},\frac{3\pi}{4})$, yielding $\epsilon_{1}=\gamma_2$. Lastly, the local minimums of the $\epsilon_1$ landscape are given by the points $(\frac{\pi}{2},0)$, $(0,\frac{\pi}{2})$, $(\pi,\frac{\pi}{2})$, $(\frac{\pi}{2},\pi)$, satisfying 
\begin{eqnarray}
\frac{\partial \epsilon_1}{\partial \alpha_x}=\frac{\partial \epsilon_1}{\partial \alpha_y}=0, \frac{\partial^2 \epsilon_1}{\partial \alpha_x ^2}>0,\det \mathcal{H}>0,
\end{eqnarray}
and yielding $\epsilon_1=1-\gamma_1$. Similar analysis can also be done for $\epsilon_2$, such that 
\begin{eqnarray}
\underset{\alpha_x,\alpha_y}{\max} \epsilon_1 &=& \gamma_1,\nonumber  \\ 
\underset{\alpha_x,\alpha_y}{\max} \epsilon_2 &=& \gamma_2 <\gamma_1,\nonumber\\
\underset{\alpha_x,\alpha_y}{\max} 1-\epsilon_1 &=& \gamma_1,\nonumber \\ 
\underset{\alpha_x,\alpha_y}{\max} 1-\epsilon_2 &=& \gamma_2<\gamma_1,
\end{eqnarray}
and $(1- \gamma_1)<(1-\gamma_2)<\gamma_2<\gamma_1$. Hence it is proved that among all eight eigenvalues obtained from the single party density matrices, $\gamma_1$ is the maximum over the allowed ranges of $\alpha_x,\alpha_y$, with $\alpha_z$ being fixed and under the conditions $\gamma_{1} \geq \delta_{1}$ and $\gamma_1 \geq \gamma_{2} \geq  \delta_{2}$. Therefore, the GGM in this case reduces to
\begin{eqnarray}
    \mathcal{G}=1-\gamma_1=\min\{\mathcal{G}_1,\mathcal{G}_2\}.
\end{eqnarray}

In situations where $\mathcal{G}$, $\mathcal{G}_1$, and $\mathcal{G}_2$ are not obtained from the eigenvalues of a single-qubit density matrix, one needs to investigate all possible bipartitions of respectively  $\ket{\Psi}_N$, $\ket{\Phi}_{N_1}$, $\ket{\Phi}_{N_2}$, and  the corresponding eigenvalues obtained from appropriate reduced density matrices. The  dependence of these eigenvalues on the state parameters as well as the parameters of the unitary operators makes analytical investigation of the GGMs difficult. Let us estimate here how the numerical complexity grows when one wants to perform ECP over \(N\) unit cells. For achieving this, one needs to operate $N-1$ entangling unitaries, i.e., the number of optimization parameters is $3(N-1)$. Also, for $N$ number of  $m$-qubit unit cell, the calculation of GGM demands eigenavalues of  $2^{{Nm}-1}-1$ number of reduced density matrices and the corresponding maximum eigenvalues. 
We perform optimization of the GGM function over the real $(\alpha_x,\alpha_y,\alpha_z)$ parameter space using a derivative-free optimization algorithm, where we sample the initial points of this search via a random global search by choosing $\alpha_j$s from random uniform distribution in the range between $0$ and $\frac{\pi}{2}$. See Appendix~\ref{sec:ggm_proposition_numerical} for a detailed discussion on the numerical analysis. Our extensive numerical analysis involving randomly generated quantum states of small to moderately high number of parties ($N\leq 6$) suggests that for specific values of the parameters defining the unit states, a unitary operator $\mathcal{U}_2$ can always be designed such that $\mathcal{G}=\min\{\mathcal{G}_1,\mathcal{G}_2\}$. 
Assuming this to be true for quantum states with arbitrary number of qubits, the condition in Proposition I can be relaxed, and Proposition II can be proposed.

\begin{figure}
    \centering
    \includegraphics[width=\linewidth]{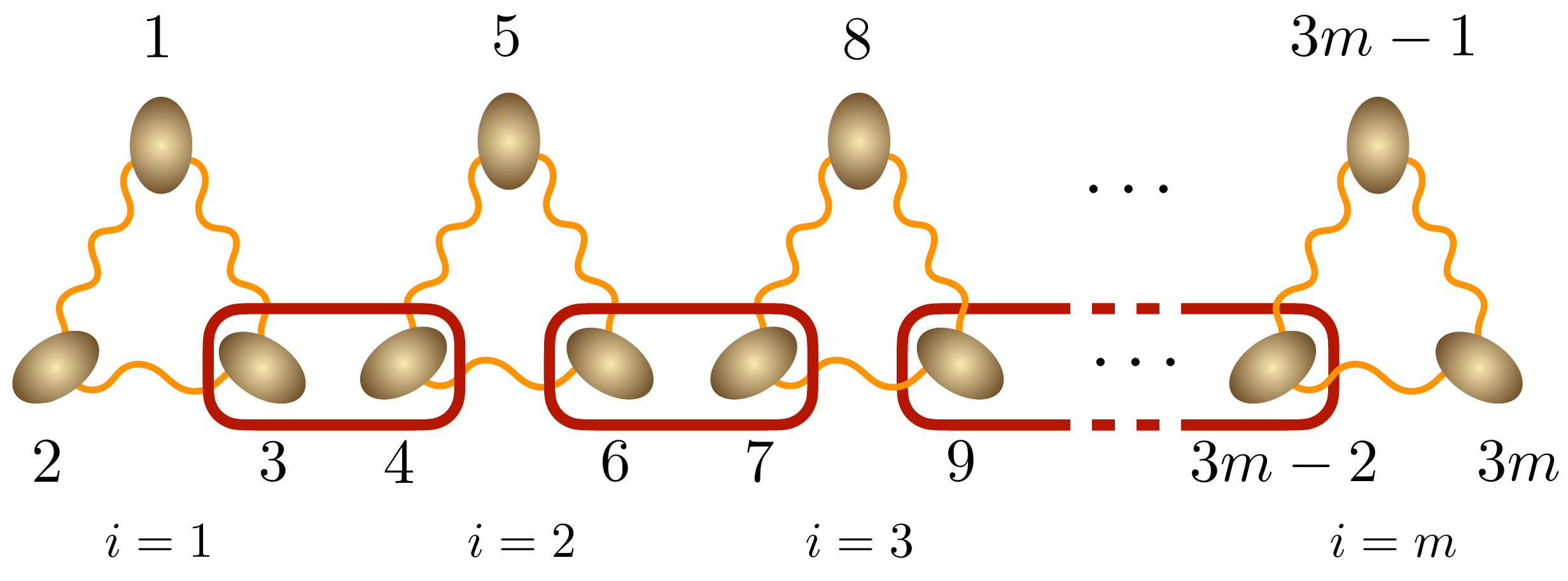}
    \caption{(Color online.) Building a \emph{linear} network with unit cells constituted of three qubits described by the state $\ket{\Phi}$, via application of two-qubit unitary operators of the form $\mathcal{U}_2$ on qubit-pairs shared by two consecutive unit states.}
    \label{fig:schem2}
\end{figure}

\noindent\textbf{$\blacksquare$ Proposition II.} \emph{For two arbitrary pure unit states $\ket{\Phi}_{N_1}$ and $\ket{\Phi}_{N_2}$, an optimal unitary operator $\mathcal{U}_2$ can always be designed such that the resulting state $\ket{\Psi}_N$, obtained by operating $\mathcal{U}_2$ on a pair of qubits constituted with one qubit from each of the unit states, gives the maximal  GGM given by  
\begin{eqnarray}
\max_{\{\mathcal{U}_2\}} \mathcal{G}=\min\{\mathcal{G}_1,\mathcal{G}_2\},    
\end{eqnarray}
where the maximization is performed over the set of parameters in \(\mathcal{U}_2\) to maximize the GGM in the output state. }\\

\noindent Proposition II can be recursively used to create multi-qubit genuinely multiparty entangled states starting from more than two arbitrary multi-qubit pure unit states, thereby establishing a connection between the GGM for the output and the input states. 

\noindent\textbf{$\blacksquare$ Proposition III.} \emph{For $m$ arbitrary pure unit states $\{\ket{\Phi}_{N_i}\}$ having GGMs, $\{\mathcal{G}_i\}$, $i=1,2,\cdots, m$, with  $N_i\geq 2$ and $\mathcal{G}_i>0$ $\forall i$, a set of $m-1$ two-qubit unitary operators $\{\mathcal{U}_2^j,j=1,2,\cdots,m-1\}$ can be constructed  such that the resulting state
\begin{eqnarray}
\ket{\Psi}_N=\bigotimes_{j=1}^{m-1}\mathcal{U}_2^{j}\ket{\Phi}_N,
\end{eqnarray}
obtained by operating the two-qubit unitary operators $\{\mathcal{U}_2^j\}$ on $m-1$ pairs of qubits with each pair constituted with two qubits from two different unit states, is genuinely multiparty entangled with the maximum GGM,   
\begin{eqnarray}
\max_{\{\mathcal{U}_2^j\}}\mathcal{G}=\min_i\{\mathcal{G}_i\}.    
\end{eqnarray}
}

Let us now stress some of the important points about these Propositions.
\begin{enumerate}
    \item[\textbf{P1.}] Proposition III implies that the resulting multi-qubit state $\ket{\Psi}_N$ will always be genuinely multiparty entangled as long as both $\mathcal{G}_i>0$ $\forall i$. Note, however, that if the initial unit state, \(N_i\), is \(k\)-separable having vanishing GGM, as mentioned before, it is again possible to apply a two-qubit unitary operator to first produce a \(N_i\)-party  state with \(\mathcal{G}_i>0\). 
    
    \item[\textbf{P2.}] Proposition III requires neither the $m-1$ two-qubit unitary operators used to create the $N$-qubit state, nor the $m$ unit states to be identical to each other. However, in terms of resource, it is indeed useful to be able to create large quantum networks using identical unitary operators, or only one type of unit states with fixed number of qubits. We shall explore the occurrence of these situations in subsequent Sections. 
    
    \item[\textbf{P3.}] Note here that the optimal unitary operator $\mathcal{U}_2$ for joining two specific multi-qubit states is turned out to  be not unique. This non-uniqueness of $\mathcal{U}_2$ for fixed pair of unit states is a crucial point which we shall elaborate in the next Section. 
    
    \item[\textbf{P4.}] Note also that the above Propositions does not include the situation where one intends to merge a single-qubit state with a multi-qubit state. However, large multiparty state can also be created by adding one auxiliary qubit at a time with a multi-qubit state of $N\geq 2$.  In the next sections, we shall point out that such a construction is rather special, and discuss its performance. 
    
    \item[\textbf{P5.}] Let us point out here that  in our numerical search, both Haar-uniformly sampled states as well 
    as states like the generalized GHZ and the generalized W states that constitute sets of measure zero are separately taken as unit states. In all these cases, our assumption that the GGM is governed by the single-qubit reduced density matrices are found to be valid. It indicates that Proposition II potentially holds for a wide range of states.
    \item[\textbf{P6.}] It is important to note that applying the same unitary operator at different steps of the ECP requires the same values of the control parameters $(\alpha_x, \alpha_y, \alpha_z)$. However,  there is no intuition to assume that the optimization of the GGM would lead to the same values of these parameters. Therefore, we expect $U_d$ to be controlled by different values of $(\alpha_x, \alpha_y, \alpha_z)$ for different steps of the protocol, while the form of $U_d$ being the same in every step.
\end{enumerate}

\begin{figure*}
\includegraphics[width=\textwidth]{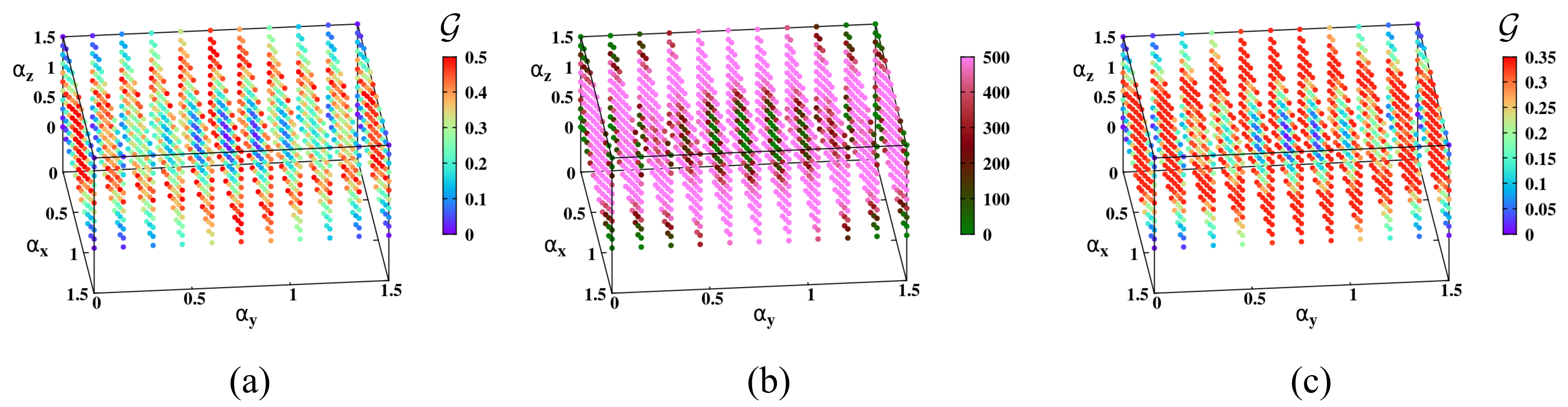}
\caption{(Color online.) \textbf{Non-uniqueness of unitary operators.} (a) The set $\mathcal{S}_U$ corresponding to two copies of three-qubit GHZ states is demonstrated on different $(\alpha_x,\alpha_y, \alpha_z)$-space in the range $0\leq \alpha_{j}\leq\frac{\pi}{2}$, $j=x,y,z$. The colors of different points $(\alpha_x,\alpha_y,\alpha_z)$ signify the values of GGM of the output six-qubit state joined by the unitary operator corresponding to that point. (b) The number of output states whose  GGMs get maximized at the specific triplets \((\alpha_x, \alpha_y, \alpha_z)\) belonging to the set $\mathcal{S}_U$ is depicted for $500$ pairs of randomly chosen three-qubit states. The figure (c) corresponds to results similar to (a), where a GHZ and a W state are joined. See Sec.~\ref{subsec:specific_cases}. All the axes are dimensionless. }
\label{fig:alpha_space}
\end{figure*}

\section{Building networks with three-qubit unit states}
\label{sec:three_qubit_units}

To build a genuinely multiparty entangled quantum states using \emph{unit states of fixed number of qubits}, we consider three-qubit genuinely multiparty entangled unit states and two-qubit unitary operators as resource.  We start with \emph{identical} three-qubit unit states of the form
\begin{eqnarray}
\ket{\Phi}_3 &=& \sum_{i=1}^8a_{i}\ket{b_i},
\label{GHZboom}
\end{eqnarray}
where $\{a_{i} \in \mathbb{C}\;\forall i\}$, and $\{\ket{b_i}\}$ is the product basis for three-qubits constituted of the single-qubit computational basis. To simplify the notation, we skip the subscript \(3\) from \(\ket{\Phi}_3\).  It is convenient to write $\ket{\Phi}$ as    
\begin{eqnarray}
\ket{\Phi} &=& \ket{A}\ket{0}+\ket{B}\ket{1}\nonumber \\
&=&\ket{0}\ket{E}+\ket{1}\ket{F},
\label{GHZkaboom}
\end{eqnarray}
where
\begin{eqnarray}
\ket{A} &=& a_{1}\ket{00}+a_{2}\ket{01}+a_{4}\ket{10}+a_{7}\ket{11},\nonumber\\
\ket{B} &=& a_{3}\ket{00}+a_{5}\ket{01}+a_{6}\ket{10}+a_{8}\ket{11},\nonumber\\
\ket{E} &=& a_{1}\ket{00}+a_{3}\ket{01}+a_{2}\ket{10}+a_{5}\ket{11},\nonumber\\
\ket{F} &=& a_{4}\ket{00}+a_{6}\ket{01}+a_{7}\ket{10}+a_{8}\ket{11}.
\end{eqnarray}
Let us consider the initial state of the system to be made of $m$ disconnected identical three-qubit pure unit states,  given by $\ket{\Phi}_{3m}=\ket{\Phi}^{\otimes m}$. In order to create the $3m$-qubit pure state, $\ket{\Psi}_{3m}$ (see Sec.~\ref{sec:general_protocol}), $m-1$ two-qubit unitary operators $\{\mathcal{U}_2^j,j=1,2,\cdots,m-1\}$ is applied on $m-1$ pairs of qubits, such that each pair is consisting of one qubit from two different unit states (see Fig.~\ref{fig:schem2} for the specific labels used for the qubits in different unit cells). We now present the recursion relation to obtain $\ket{\Psi}_{3m}$ (see Appendix~\ref{sec:recursion_derivation} for a derivation). 

\noindent\textbf{$\blacksquare$ Proposition IV:} \emph{Applying the two-qubit unitary operator, $\mathcal{U}_2$, $m-1$ times on the initial state, $\ket{\Phi}_{3m}=\ket{\Phi}^{\otimes m}$, a $3m$-qubit state of the form
\begin{equation}
\ket{\Psi}_{3m} = \ket{X}^{m-1}\ket{E}+\ket{Y}^{m-1}\ket{F}
\label{eq:resultant_state}
\end{equation}
is obtained, with} 
\begin{widetext}
\begin{eqnarray}
\ket{X}^{m-1} &=&\left[\ket{X}^{m-2}(a_{1}\ket{0}+a_{2}\ket{1})+\ket{Y}^{m-2}(a_{4}\ket{0}+a_{7}\ket{1})\right]U_d^{m-1}\ket{00} \nonumber\\
&&+\left[\ket{X}^{m-2}(a_{3}\ket{0}+a_{5}\ket{1})+\ket{Y}^{m-2}(a_{6}\ket{0}+a_{8}\ket{1})\right]U_d^{m-1}\ket{10},\nonumber\\
\ket{Y}^{m-1} &=&\left[\ket{X}^{m-2}(a_{1}\ket{0}+a_{2}\ket{1})+\ket{Y}^{m-2}(a_{4}\ket{0}+a_{7}\ket{1})\right]U_d^{m-1}\ket{01}\nonumber\\ 
&&+\left[\ket{X}^{m-2}(a_{3}\ket{0}+a_{5}\ket{1})+\ket{Y}^{m-2}(a_{6}\ket{0}+a_{8}\ket{1})\right]U_d^{m-1}\ket{11}, 
\end{eqnarray}
\end{widetext}
\emph{where $\ket{X}^{m-1}$ and $\ket{Y}^{m-1}$ can be obtained for arbitrary $m$ starting from  
\begin{eqnarray}
\ket{X}^1 &=& \ket{A}U_d^1\ket{00}+\ket{B}U_d^1\ket{10},\nonumber \\
\ket{Y}^1 &=& \ket{B}U_d^1\ket{11}+\ket{A}U_d^1\ket{01}. 
\end{eqnarray}}

\noindent It is clear from \textbf{Proposition III} that $\ket{\Psi}_{3m}$ is genuinely multiparty entangled provided the initial resource state, $\ket{\Phi}$, is genuinely multiparty entangled. Note here that identical unit states is an idealized scenario where no error in the preparation of three-qubit unit states is assumed. In reality, however, the unit states may differ from each other due to imperfect preparation and 
similar procedure can be opted for obtaining 
a recursion relation for different three-qubit unit states although the relation for the output state is  much more involved.  

\subsection{Nonuniqueness of unitaries}
\label{subsec:specific_cases}

We will now discuss the set of optimal unitary operators which lead to six-qubit output states, starting from the three-qubit initial states. 
Note here that the three-qubit unit states may belong to both the GHZ- and the W-class \cite{Durvidal2002}, so that different scenarios involving (a) two GHZ-class states, (b) two W-class states, and (c) a combination of GHZ- and W-class states  can be considered. In the following, we demonstrate the behavior of  GGM of the resulting states obtained from scenarios, (a), (b) and (c).

\subsubsection{Optimal unitaries for merging two  GHZ-class states} 
\label{subsubsec:ghz-ghz}

Let us describe a set of optimal unitary operators, $\mathcal{S}_U = \{\mathcal{U}_2\}$, which can generate a six-qubit state, $\ket{\Psi}$, having maximal GGM, i.e., \(0.5\), from  two copies of the initial GHZ states, given by  
\(\ket{\Phi}_{GHZ}=\frac{1}{\sqrt{2}}(\ket{000}+\ket{111})\).
Determination of the six-qubit state using Eq.~(\ref{eq:resultant_state}), and subsequent computation of the eigenvalues of the reduced density matrices for different bipartitions of the state indicate that the GGM of the resultant state  in terms of parameters of \(U_2\), i.e., \(\alpha_j, j=x,y,z\) reduces to
\begin{eqnarray} \mathcal{G}=1-\max\{\lambda_1^1,\lambda_3^{123},\lambda_3^{124}\}
\end{eqnarray}
with $\lambda_1^1=\lambda_1^2=\lambda_1^5=\lambda_1^6=\frac{1}{2},$  $\lambda_3^{123}=\frac{1}{4}[1+\cos2\alpha_y \cos2\alpha_z+\cos 2\alpha_x (\cos 2\alpha_y + \cos 2\alpha_z)]$, and $\lambda_3^{124}=\frac{1}{4}[1+\sin2\alpha_y \sin2\alpha_z+\sin 2\alpha_x (\sin 2\alpha_y + \sin 2\alpha_z)]$, originating from only single- and three-qubit reduced density matrices. A large number of convenient choices of the parameters $\alpha_j,j=x,y,z$, is possible providing a set $\mathcal{S}_U$ of non-unique two-qubit unitary operators of the form $U_d$, ensuring that the GGM of the resulting state is (as per Proposition III) \(\mathcal{G}=\min\{\mathcal{G}_1,\mathcal{G}_2\}=\mathcal{G}(\ket{\Phi})=1/2\).

In order to see how the values of the GGM of $\ket{\Psi}$ varies with the parameters of $U_d$,  we plot the GGM as a function of $\{\alpha_j,j=x,y,z\}$ in Fig.~\ref{fig:alpha_space}(a). This figure clearly indicates that a high value of GGM is favourable if any one of the unitary parameters $\{\alpha_j\}$ vanishes. It also shows that the number of unitary operators that can produce a six-qubit genuinely multiparty entangled state with the maximum GGM is very small (approximately $4.7\%$ unitaries belong to the set \(\mathcal{S}_U\) among the total number of $U_d$ generated which is \(3.2 \times 10^4\)  ).  

We will now establish that such a set of unitary operators which maximizes the GGM of the resulting state exists irrespective of the initial states. 
For this analysis, we Haar uniformly generate two arbitrary  three-qubit states, denoted by $\ket{\Phi}_c^1$ and $\ket{\Phi}_c^2$ which eventually belong to the GHZ-class. We randomly generate a large number of such pairs of states, and for each such pair, a large set $\mathcal{S}_U$ consisting of two-qubit unitary operators $\mathcal{U}_2$ is found, such that for each $\mathcal{U}_2\in\mathcal{S}_U$, the GGM of the resulting state is given by $\mathcal{G}=\min\{\mathcal{G}_1,\mathcal{G}_2\}$, $\mathcal{G}_1$ ($\mathcal{G}_2$) being the GGM of $\ket{\Phi}_c^1$ ($\ket{\Phi}_c^2$) (see also Appendix~\ref{sec:ggm_proposition_numerical}). 

\textbf{Observation.} It is interesting to observe that there is a large overlap between these sets $\mathcal{S}_U$ corresponding to different pairs of ($\ket{\Phi}_c^1, \ket{\Phi}_c^2)$, implying the existence of two-qubit unitary operators that can combine pairs of large number of randomly generated three-qubit states from the GHZ-class, so that the Proposition III remains valid for the resulting states. For ease of reference, we call these unitary operators as \emph{universal} unitary operators. Secondly, this observation is useful in terms of resource minimization while creating large quantum networks using three-qubit unit states (see point \textbf{P2} in Sec.~\ref{subsec:protocol}). We demonstrate this in Fig.~\ref{fig:alpha_space}(b), where we count the number of output states whose GGMs get maximized for specific values of \(\alpha_x, \alpha_y\), and  \(\alpha_z\) belonging to the set \(\mathcal{S}_U\). The analysis is performed by generating $5\times 10^2$ random three-qubit states,  $\ket{\Phi}_c$.

\subsubsection{Merging a GHZ and a W state}
\label{subsubsec:ghz-w}

Unlike identical copies, if the initial state of the six-qubit system is given by the product of a GHZ state, $\ket{\Phi}_{GHZ}$  and a $W$ state given by 
\(\ket{\Phi}_W=\frac{1}{\sqrt{3}}(\ket{001}+\ket{010}+\ket{100})\),    
similar observations as discussed in the case of two GHZ-class states (see Sec.~\ref{subsubsec:ghz-ghz}) emerge. The GGM of the resultant six-qubit state, given by 
\begin{eqnarray}
\mathcal{G}=\min\{\mathcal{G}(\ket{\Phi}_{GHZ}),\mathcal{G}(\ket{\Phi}_{W})\}=\mathcal{G}(\ket{\Phi}_{W})=\frac{1}{3}, 
\end{eqnarray}
follows Proposition III, while the eigenvalues contributing in the computation of $\mathcal{G}$ 
\begin{eqnarray}
\lambda_1^5 &=& \lambda_1^6 = \frac{2}{3},\nonumber \\
\lambda_{3}^{123} &=& \frac{1}{48}(12+12\cos{2\alpha_x}\cos{2\alpha_y}+\sqrt{2A}),\nonumber \\ 
\lambda_{3}^{124} &=& \frac{1}{48}(12+12\sin{2\alpha_x}\sin{2\alpha_y}+\sqrt{2B}),
\end{eqnarray}
with  
\begin{widetext}
\begin{eqnarray}
A &=& 42+40(\cos{2(\alpha_x-\alpha_y)}+\cos{2(\alpha_x+\alpha_y)})+\cos{4(\alpha_x-\alpha_y)}+\cos{4(\alpha_x+\alpha_y)}+18 (\cos{4\alpha_x} +\cos{4\alpha_y})\nonumber\\
&&+32(\cos{2\alpha_x}+\cos{2\alpha_y})^2\cos{4\alpha_z}, \\ 
B &=& 42+40(\cos{2(\alpha_x-\alpha_y)}-\cos{2(\alpha_x+\alpha_y)})+\cos{4(\alpha_x-\alpha_y)}+\cos{4(\alpha_x+\alpha_y)}-18 (\cos{4\alpha_x} +\cos{4\alpha_y})\nonumber\\
&&-32(\sin{2\alpha_x}+\sin{2\alpha_y})^2\cos{4\alpha_z}, 
\end{eqnarray}
\end{widetext}
which are obtained by diagonalizing the reduced states corresponding to qubit $5$ (or qubit $6$), qubits $123$, and qubits $124$ of the six-qubit system respectively.
Similar numerical analysis again reveals that a larger volume of the $\alpha_j$-space ($j=x,y,z$) (i.e., approximately \(41.3\%\) of the generated set of unitary operators, \(3.2 \times 10^4\), leading to maximum \(\mathcal{G}\)), compared to the merging of two copies of the GHZ states, correspond to $\mathcal{S}_U$ when a GHZ and a W states are combined (see Fig.~\ref{fig:alpha_space}(c)).

\subsection{Optimal distribution of resource}
\label{subsec:optimal_resource_distribution}

In order to create a genuinely multi-qubit entangled state $\ket{\Psi}_N$ of $N$-qubits, multiple choices for the set of values $\{N_1,N_2,\cdots,N_m\}$ denoting the number of qubits in the unit states are possible. However, it is not at all clear whether all these choices are equivalent, or a subset of these choices are more beneficial in order to obtain higher multiparty entanglement in the resultant state. This information can be useful in situations, where one is forced to prepare smaller multi-qubit states in the laboratory in order to create larger multi-qubit entangled state using our protocol. It can be due to the fact that  creating large multi-qubit states in certain physical substrates like photons  is difficult. We now address this issue, and demonstrate the effect in the distribution of the support of the unit states on the GGM of the final state.

\begin{figure}
	\includegraphics[width=1.0\linewidth]{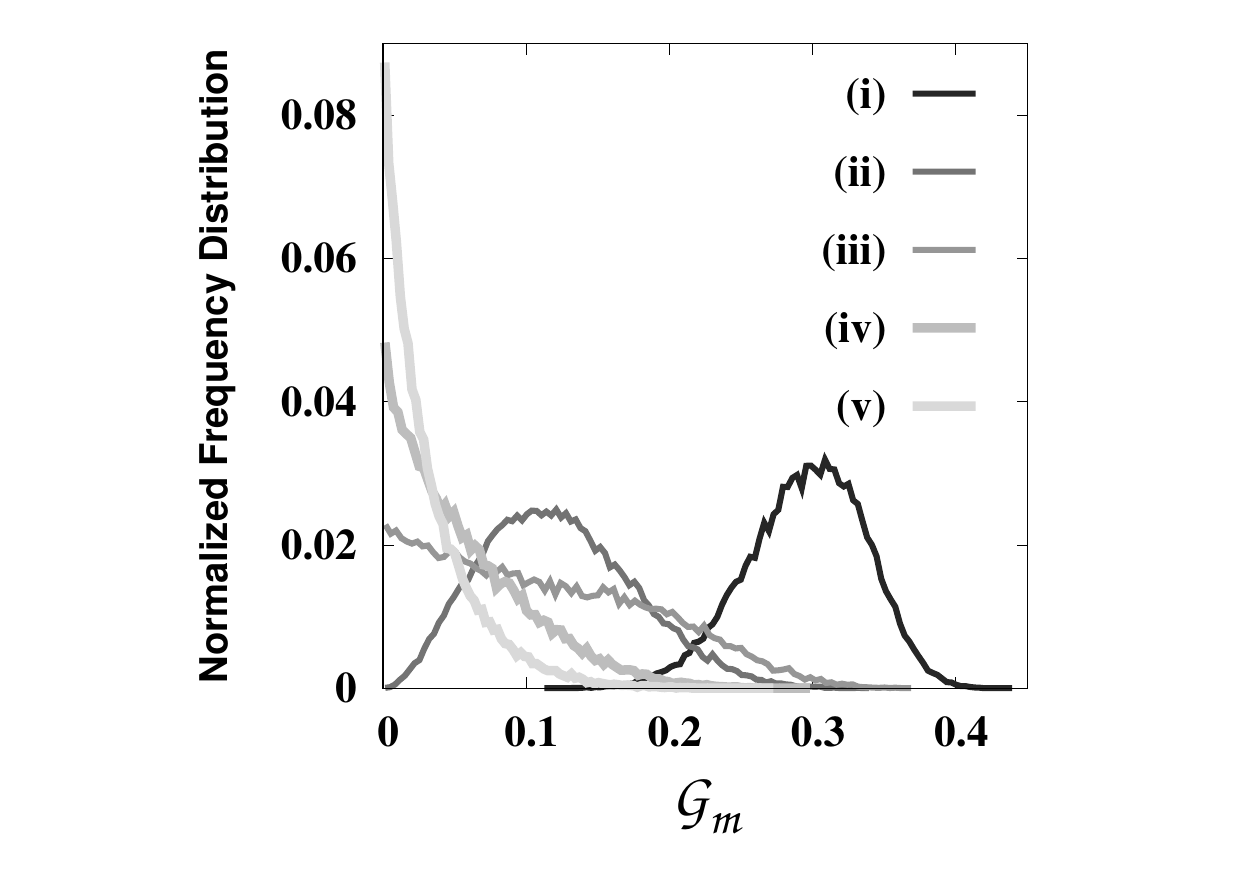}
	\caption{ (Color online.) \textbf{Optimally distributing resource states.} Normalized frequency distribution (vertical axis) against optimized  GGM, $\mathcal{G}_m$ (horizontal axis) of six-party  genuinely multiparty entangled state as output with different initial resources. Labels for different plots (i)-(v) respectively represent the unit states as  $(\ket{\Phi}_5, \ket{\Phi}_1)$, $(\ket{\Phi}_{3,GHZ}, \ket{\Phi}_{3, GHZ})$, $(\ket{\Phi}_4, \ket{\Phi}_2)$, $(\ket{\Phi}_{3, GHZ}, \ket{\Phi}_{3, W})$, $(\ket{\Phi}_{3,W}, \ket{\Phi}_{3,W})$
	(see Table.~\ref{table:resource_distribution}).
	In each case, we create $5\cross10^4$  output states from Haar-uniformly generated initial resource states. Both the axes are dimensionless.}
	\label{fig:resource_distribution}
\end{figure}

For the purpose of demonstration, we consider the case of $N=6$, which can be obtained from different unit cells of sizes (a) $(N_1=3,N_2=3)$, (b) $(N_1=4,N_2=2)$, and (c) $(N_1=5,N_2=1)$.  For each combinations of $(N_1,N_2)$, we Haar uniformly generate a large number of quantum states $\ket{\Phi}_{N_1}$ and $\ket{\Phi}_{N_2}$. For each pair of such unit states $\{\ket{\Phi}_{N_1},\ket{\Phi}_{N_2}\}$, we apply a two-qubit unitary operator $\mathcal{U}_2$ on a pair of qubits shared by the two unit states, and determine the maximum GGM of the resultant state over the set $\mathcal{S}_U$, as
\begin{eqnarray}
\mathcal{G}_{m}=\underset{\mathcal{S}_U}{\max} \;\{\mathcal{G}\}.
\end{eqnarray}
In the case of two non-identical three-qubit states, we consider three specific scenarios -- (i) a pair of different unit states both belonging to the GHZ-class,  (ii) a pair of different unit states both belonging to the W class, and (iii) a pair of unit states, one from the GHZ-class and the other from the W class.  Fig.~\ref{fig:resource_distribution} depicts the normalized frequency distribution of $\mathcal{G}_m$ in all of these scenarios, where a set of  \(5 \times 10^4\) Haar-uniformly generated  unit states are used in each cases.

\begin{table} 
\begin{tabular}{|c|c|c|c|}
\hline
          \begin{tabular}{c|c}
          No. & Types of unit states \\
          \hline
        (i)  & $\ket{\Phi}_5$, $\ket{\Phi}_1$ \\
          \hline
        (ii) & $\ket{\Phi}_{3,GHZ}$, $\ket{\Phi}_{3,GHZ}$  \\
          \hline
        (iii)& $\ket{\Phi}_4$, $\ket{\Phi}_2$ \\
          \hline
        (iv) & $\ket{\Phi}_{3,GHZ}$, $\ket{\Phi}_{3,W}$ \\
          \hline
         (v) & $\ket{\Phi}_{3,W}$, $\ket{\Phi}_{3,W}$ \\
          \end{tabular}
          &
          \begin{tabular}{c|c}
            $\langle\mathcal{G}_m\rangle$ & $\sigma_{\mathcal{G}_m}$  \\ 
          \hline 
           $0.295$  &$0.041$  \\
          \hline
         $0.122$  &$0.052$    \\
         \hline 
           $0.111$  &$0.076$  \\
          \hline
         $0.056$  & $0.046$  \\
         \hline 
           $0.033$  &$0.032$  \\
         \end{tabular}   \\      
\hline                      
\end{tabular}
\caption{ Mean and standard deviation corresponding to different resource distribution.}
\label{table:resource_distribution}
\end{table} 

It is clear from Fig.~\ref{fig:resource_distribution} that the mean of the distributions, $\langle\mathcal{G}_m\rangle$, which is tabulated in Table.~\ref{table:resource_distribution} along with the standard deviation $\sigma_{\mathcal{G}_m}$, is maximum for the case $(N_1=5,N_2=1)$, and is minimum for $(N_1=3,N_2=3)$ when both the states belong to the W class.  Also, the case of $(N_1=3,N_2=3)$ with both states coming from the GHZ-class is better than the case of $(N_1=4,N_2=2)$ for generating higher multiparty entangled states on average. Note here that the situation $(N_1=5,N_2=1)$ is different from the rest of the combinations of $N_1$ and $N_2$ since the Proposition III (see Sec.~\ref{subsec:protocol}) does not apply to this case. Furthermore,  $(N_1=5,N_2=1)$ is  the most expensive one according to the number of the qubits of the initial resource, since it is difficult to create genuine multi-qubit entangled states with higher number of parties. However, this bottle-neck can be resolved by noting that one can use two genuinely multi-qubit entangled states, one of three qubits and the other of two qubits, to create the five-qubit entangled state using the same protocol.  

\begin{figure}
\includegraphics[width=0.7\linewidth]{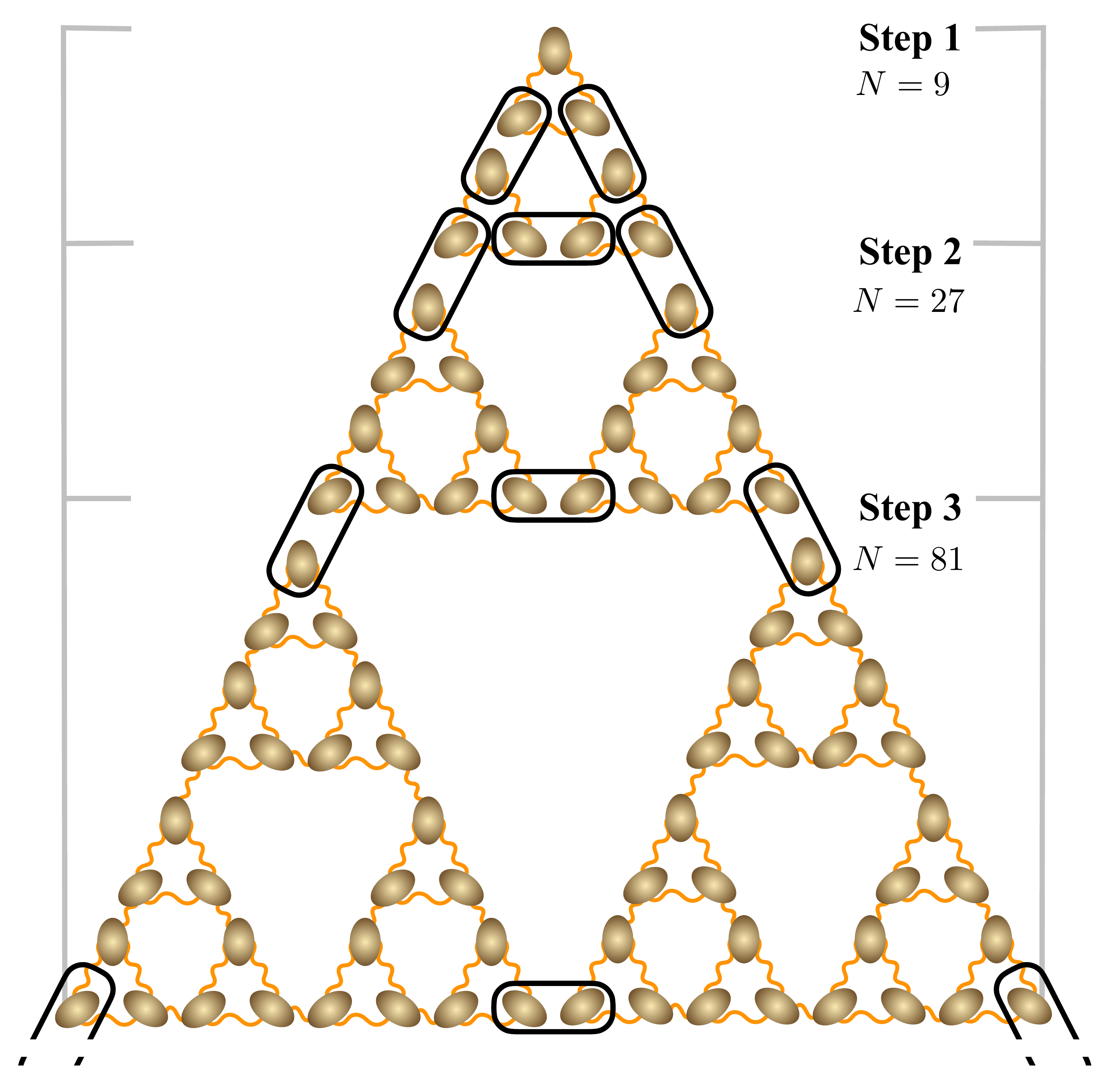}
\caption{ (Color online.) \textbf{Triangle network.} Protocol for building a \emph{triangle} network, starting from unit cells constituted of three qubits and described by the state $\ket{\Phi}$. In Step 1 (see Sec.~\ref{subsec:geometry}), three unit states are combined via two-qubit unitary operators to form a nine-qubit state. In Step 2, the nine-qubit states are used as unit states. The network upto Step 3 is shown, finally producing a state of $81$ qubits.}
\label{fig:triangle} 
\end{figure}

\subsection{Quantum networks of different geometry}
\label{subsec:geometry}

We now demonstrate how the protocol proposed and discussed over Sec.~\ref{sec:three_qubit_units} can be used to create large quantum networks of a geometry other than the linear geometry (see Fig.~\ref{fig:schem2} and the discussion in Sec.~\ref{subsec:protocol}). More specifically, we illustrate  how a quantum network of triangular geometry can be constructed starting from the three-qubit GHZ states as unit states, and by using two-qubit unitary operators. This protocol can also be modified for any three-qubit unit states, although the computation of the resultant state will be more cumbersome. 

The steps of the method are given as follows.  
\begin{enumerate}
    \item In the first step, take three  GHZ states and apply two-qubit unitary operators of the form $\mathcal{U}_2$ on three pairs of qubits (see Fig.~\ref{fig:triangle}), such that each pair is shared by two different GHZ states. This gives rise to a nine-qubit genuinely multiparty entangled state $\ket{\Psi}^1$ given by
    \begin{eqnarray}
    \ket{\Psi}^1=\mathcal{U}_2^3\mathcal{U}_2^2\mathcal{U}_2^1\ket{\Phi}^0=\mathcal{U}\ket{\Phi}^0,
    \end{eqnarray}
    Here, $\ket{\Phi}^0=\ket{\Phi}^{\otimes 3}$, $\ket{\Phi}$ is a three-qubit GHZ state, $\mathcal{U}_2^j$($j=1,2,3$) are the three unitary operators applied to three different pairs of qubits, $\mathcal{U}=\mathcal{U}_2^3\mathcal{U}_2^2\mathcal{U}_2^1$, and the superscripts to $\ket{\Phi}$ and $\ket{\Psi}$ denote the step of the protocol. Note here that the form of $\ket{\Psi}^1$ would be different from the $\ket{\Psi}$ obtained by joining three GHZ states using two unitary operators, as described in Sec.~\ref{sec:three_qubit_units}, and when simplified, it takes the form
    \begin{eqnarray}
     \ket{\Psi}^1 &=& \frac{1}{2\sqrt{2}}\big[\ket{000}\mathcal{U}\ket{\xi_0}+\ket{001}\mathcal{U}\ket{\xi_1} \nonumber \\ &&+\ket{010}\mathcal{U}\ket{\xi_2} +\ket{011}\mathcal{U}\ket{\xi_3}+\ket{100}\mathcal{U}\ket{\xi_4}\nonumber \\
     &&+\ket{101}\mathcal{U}\ket{\xi_5}+\ket{110}\mathcal{U}\ket{\xi_6}+\ket{111}\mathcal{U}\ket{\xi_7}\big],
    \end{eqnarray}
    where the states $\ket{\xi_i}$,$i=0,\cdots,7$, correspond to the qubits on which the two-qubit unitaries act, and are given by
    \begin{eqnarray}
    \ket{\xi_0} &=& \ket{000000}, \ket{\xi_4}=\ket{111111},\nonumber \\
    \ket{\xi_1} &=& \ket{010001},\ket{\xi_5}=\ket{101110},\nonumber\\ 
    \ket{\xi_2} &=& \ket{000110},\ket{\xi_6}=\ket{111001},\nonumber\\
    \ket{\xi_3} &=& \ket{010111},\ket{\xi_7}=\ket{101000}. 
    \end{eqnarray}

    \item In the next step, take $\ket{\Psi}^1$ as the unit state, and merge three of them using three unitary operators, similar to step 1 (see Fig.~\ref{fig:triangle}). 
    
    \item Continue step 2, every time replacing the unit state with the multi-qubit state obtained in the previous step.
\end{enumerate}
\textbf{Remark.} At the end of the $k$th step of the protocol, one is able to create a $3^{k+1}$-qubit state using only $3k$ two-qubit unitary operators, implying that the size of the network in terms of number of qubits grows exponentially with steps of the protocol, while the number of unitary operators required to carry out these steps grow only linearly with the number of steps. Note, however, that such a relation between number of qubits and unitary operators depend on the geometrical structure of the network.

\section{Entanglement Circulation using Many-Body Interactions: Order vs. Disorder}
\label{sec:many_body}

\begin{figure*}
\includegraphics[width=0.7\textwidth]{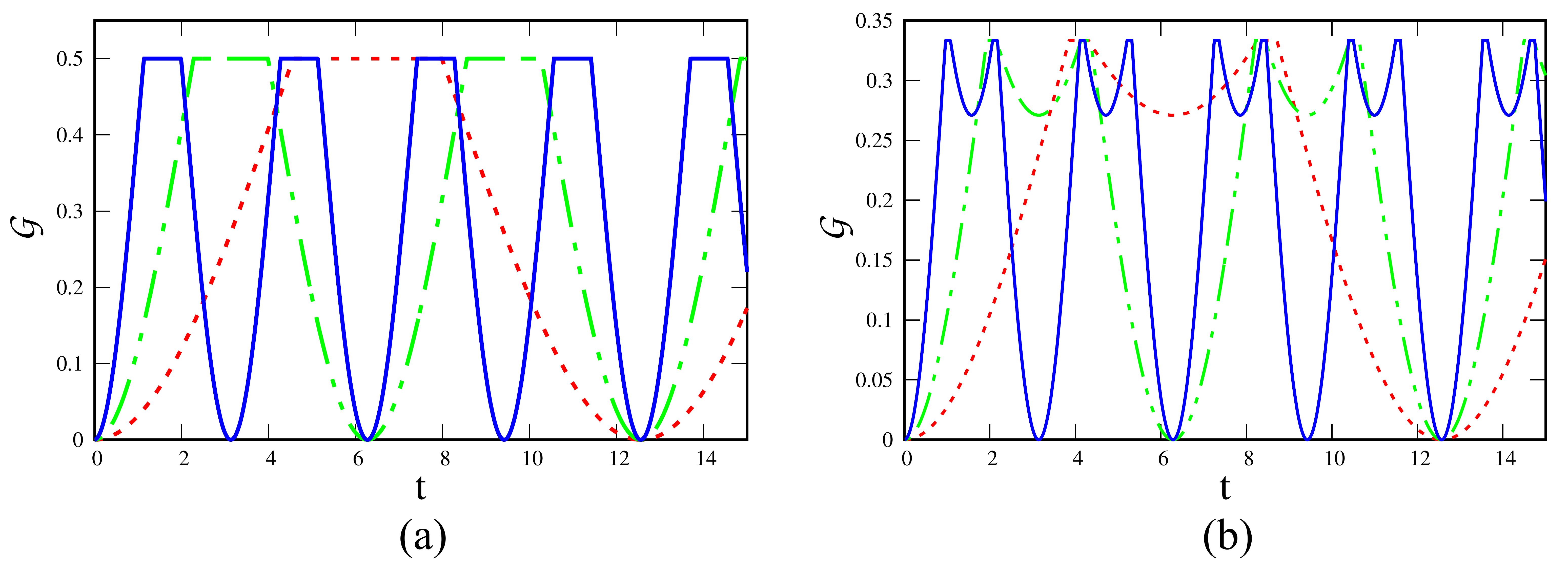}
\caption{(Color online.) \textbf{Dynamics of GGM.} Patterns of GGM (ordinate) of the six-qubit evolved  state as a function of $t$ (abscissa), with (a) two GHZ states and (b) two W states as resource. The evolution happens according to  the Hamiltonian $H_{kl}$ with $\gamma=0$, and $\Delta=0$. Different lines indicate different values of interaction strengths $J$ (solid blue line: $J=2.0$, dot-dashed green line: $J=1.0$, and dashed red line: $J=0.5$). Both the axes are dimensionless.  }
\label{fig:many_body} 
\end{figure*}

We now consider the growth of  genuinely multiparty entangled states  in networks via unitary operators emerging from many-body interactions between the qubits and determine the optimal time which leads to maximum GGM for a given Hamiltonian. The motivation lies in the fact that entangling unitary operations such as the controlled NOT and the controlled phase gates can be implemented via evolving the system using spin-spin interaction Hamiltonians~\cite{hein2006}. It is, therefore, logical to explore the possibility of implementing the ECP in a similar fashion, which we address here. The specific Hamiltonian considered in this Section is either ordered or disordered. The ordered case can be considered as the ideal situation while the evolution according to the disordered model incorporates the imperfections in the operations.

\subsection{Creating  output states via ordered spin models}
\label{subsec:orderedmodel}

Let us first present the prescription of the scheme for generating genuine multipartite entangled states when the evolution of the system is governed by the spin model without disorder.

\begin{enumerate}
    \item \emph{Preparation:} Prepare \(m\) number of \(N_i\)-qubit unit states, \(\ket{\Phi}_{N_i}\) with \(N= \sum_i N_i\), so that the initial \(N\)-qubit state can be represented as $\ket{\Phi (0)}=\ket{\Phi}_{N_1}\otimes \ldots \otimes \ket{\Phi}_{N_m}$ at initial time, $t=0$. We assume that they are genuinely multiparty entangled. 
    
    \item \emph{Evolution:} \(m'\) number of two-qubit quantum spin Hamiltonian $H_{kl}^{(r)}$ (\(r =1, \ldots, m'\)) involving spin-exchange interactions between the qubits $k$ and $l$, belonging to two different unit cells is turned on at $t>0$, such that the state $\ket{\Phi (0)} $ evolves as
    \begin{eqnarray}
        \ket{\Psi(t)}= \bigotimes_{r} \text{e}^{-\text{i}H_{kl}^{(r)}t}\ket{\Phi (0)},
    \end{eqnarray}
    where the values of the spin-exchange interaction strengths in $H_{kl}^{(r)}$ is tuned for optimal time interval to obtain the desired value of the genuine multipartite entanglement in $\ket{\Psi(t)} $. 
\end{enumerate}

\textbf{Note.} As mentioned in case of arbitrary unitary dynamics, if multipartite entanglement of one of the unit state is  not genuine, the second step can be applied first to make the unit cell multipartite entangled and then evolution is again performed between different unit cells.

For the purpose of demonstration, let us consider the two-qubit ordered XYZ model, given by the Hamiltonian connecting two unit cells, say, the \(r\)th one as 
\begin{equation}
    H_{kl}^{(r)} = \frac{J}{4} [(1+\gamma) \sigma_{k}^{x} \sigma_{l}^{x} +(1-\gamma) \sigma_{k}^{y} \sigma_{l}^{y} ] + \frac{\Delta J}{4} \sigma_{k}^{z} \sigma_{l}^{z},
    \label{eq:H_xyz}
\end{equation}
where $\sigma^{\mu}$ ($\mu = x, y, z$) are the Pauli matrices,  $J$ is the interaction strength between qubits $k$ and $l$ while $\gamma$ and $\Delta$ are respectively the $xy$- and the $z$-anisotropy parameters. Notice that there can be a situation where the spin-exchange interaction strength can be different for all qubit pairs in the \(N\)-qubit system.

Let us now investigate the pattern of GGM when the initial unit states are chosen to be three-qubit and a interacting  Hamiltonian, \(H_{kl}\), is applied between two unit cells. (Since the connection is made between two unit cells, we skip the superscript for simplicity.)  First, we consider  \(H_{kl}\) with $\gamma=0$ and $\Delta=0$ for manifestation, where the two-qubit computational basis is transformed as
\begin{eqnarray}
\text{e}^{-\text{i}H_{kl}t}\ket{0_k0_l} &=& \ket{0_k0_l}, \nonumber\\
\text{e}^{-\text{i}H_{kl}t}\ket{1_k0_l}&=& \cos{\frac{Jt}{2}}\ket{1_k0_l}-\text{i}\sin{\frac{Jt}{2}}\ket{0_k1_l},\nonumber\\
\text{e}^{-\text{i}H_{kl}t}\ket{0_k1_l} &=& \cos{\frac{Jt}{2}}\ket{0_k1_l}-\text{i}\sin{\frac{Jt}{2}}\ket{1_k0_l},\nonumber\\
\text{e}^{-\text{i}H_{kl}t}\ket{1_k1_l} &=& \ket{1_k1_l}.
\label{eq:basis_time_evolution}
\end{eqnarray}
Using Eq.~(\ref{eq:basis_time_evolution}) and Proposition IV, the form of the six-qubit resultant state $\ket{\Psi(t)}$ can be determined as a function of time, and the GGM of $\ket{\Psi(t)}$ can also be computed. Over the time evolution of the state, we observe that a competition between the maximum eigenvalues originating from the single-qubit and three-qubit density matrices takes place. During the time intervals where the maximum among all eigenvalues comes from the single-qubit density matrices and is a constant, GGM exhibits constancy over time. The duration in which  GGM remains constant can be tuned by controlling the value of interaction strength, $J$ (see Figs. ~\ref{fig:many_body}(a) and \ref{fig:many_body}(b) when the initial unit states are the GHZ  and  the W state respectively). From the perspective of implementation,   such control over the parameters in the Hamiltonian  can be important since certain  quantum information processing tasks require  fixed amounts of genuine multiparty entanglement as resource.  Overall, GGM exhibits a periodic behavior over time with a period of $T=\frac{2\pi}{J}$ for $\gamma,\Delta=0$.
Notice that for both the cases of GHZ and W states, Proposition III remains valid at every time instant, bounding the GGM of the resultant state via the GGM of the initial unit states. 

In presence of \(\gamma\) and \(\Delta\), the GGM of the output state after the evolution according to the Hamiltonian, \(H_{kl}\), again exhibits  periodic behavior with time having period  of the form \(2 \pi/f(J, \gamma, \Delta)\).

\begin{figure*}
    \centering
	\includegraphics[width=0.8\linewidth]{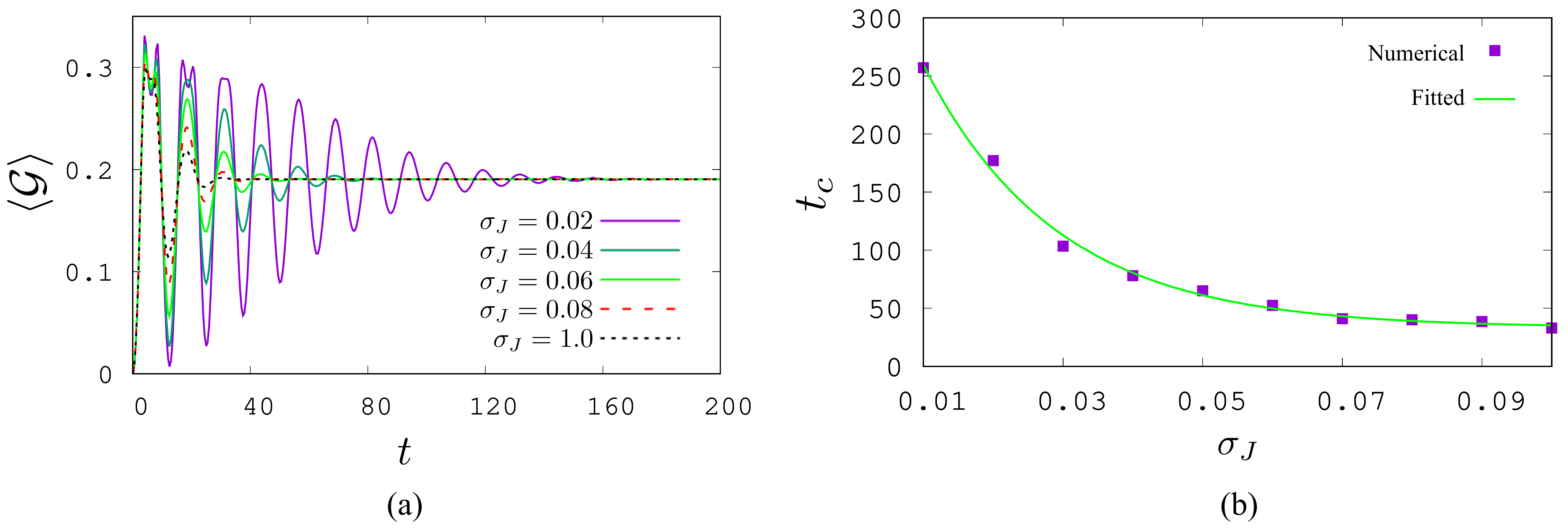}
	\caption{(Color online.) \textbf{Dynamical states via disordered systems.}  (a) Quench-averaged GGM, $\langle\mathcal{G}\rangle$ (vertical axis), of the resulting state against $time$  (horizontal axis) when two copies of the W state are chosen as the initial states. The evolution occurs according to \(H_{kl}\) with the interaction strength being chosen randomly from  Gaussian distribution with  mean $0.5$ and varying standard deviation, $\sigma_{J}=0.01$ to $\sigma_{J}=0.1$.  (b) Squares represent the critical time, $t_{c}$, above which  \(\langle \mathcal{G}\rangle\) saturates (ordinate)  with respect to $\sigma_{J}$ (abscissa) for the same initial state as in (a). Solid line is the \(\chi\)-square fit of \(t_c\) (see text for details). Both the axes are dimensionless. }
	\label{fig:dis_t_sigma}
\end{figure*}

\subsubsection{Creation of single excitation Dicke states} 

In recent times,  several proposals have been made to create the \(N\)-qubit W states in networks \cite{Tashima(3)2008, Tashima(4)2011,  zang2015, sharma2020}. We will now illustrate that  the method proposed here is also capable to deterministically produce  Dicke states with single excitation.  
To do so, let us first take a three-qubit W state, $\ket{\Phi}_W$ and single qubit auxiliary state, $\ket{0}$, so that the initial state is $\ket{\Phi}_{4}=\ket{\Phi}_W\bigotimes\ket{0}$. When the third qubit of the W state and  the auxiliary qubit evolve according to the Hamiltonian, \(H_{kl}\) with \(\gamma = \Delta =0\), the resulting state turns out to be the four-qubit  Dicke state having single excitation given by
\begin{eqnarray}
    \ket{\Psi}^{\mathcal{D}_1}_{4}=\sqrt{\frac{2}{3}}\ket{\psi^+}\ket{00}+ \sqrt{\frac{1}{6}}\ket{00}\ket{Z_1},
\end{eqnarray}
where
\begin{eqnarray}
    \ket{Z_1} &=& (e^{-\text{i}\frac{J}{2}t}\ket{\psi^+}+e^{\text{i}\frac{J}{2}t}\ket{\psi^-}),
\end{eqnarray}
with $\ket{\psi^{\pm}}=\frac{1}{\sqrt{2}}(\pm \ket{01}+\ket{10})$, $\ket{11}$, $\ket{00}$ being the eigenstates of the Hamiltonian. Notice that instead of the third qubit,  if  the Hamiltonian dynamics involves  any other qubits of the W state and the auxiliary qubit, the resulting state still remains same  due to the symmetry of the \(W\) state. Moreover, the number of excitation in the resulting state remains conserved after the evolution since the total spin angular momentum commutes with the Hamiltonian.

Let us now move further, and instead of single auxiliary qubit, let us add two auxiliary qubits, i.e.,  $\ket{\Phi}_{5}=\ket{\Phi}_W\bigotimes\ket{0}\bigotimes\ket{0}$. If the dynamics happens according to the Hamiltonian independently on the pair of qubits,  $(3, 4)$ and $(4, 5)$, the output five-qubit state reads as
\begin{eqnarray}
  \nonumber   \ket{\Psi}^{\mathcal{D}_{1}}_{5} &=& \sqrt{\frac{2}{3}}\ket{\psi^+}\ket{000}  \\
\nonumber     &+& \frac{1}{2 \sqrt{6}} \ket{00}\Big(e^{-\text{i}\frac{J}{2}t}\left\{\ket{0}\ket{Z_1}+  \sqrt{2}\ket{1}\ket{00}\right\}\\
     &+& e^{\text{i}\frac{J}{2}t}\left\{-\ket{0}\ket{Z_1}+  \sqrt{2}\ket{1}\ket{00}\right\}\Big).
\end{eqnarray}
Taking $N$ such auxiliary qubits, $\ket{0}$ and evolving  \(N\) pairs according to \(H_{kl}\),    the $(3+N)$-qubit Dicke state with single excitation can be created as  
\begin{eqnarray}
        \ket{\Psi}^{\mathcal{D}_1}_{3+N} &=& \sqrt{\frac{2}{3}}\ket{\psi^+}\ket{0}^{\bigotimes N+1}+\frac{1}{2^{\frac{2N-1}{2}}\sqrt{3}}\ket{00}\ket{Z_N},\nonumber \\ 
\end{eqnarray}
  where for $N\geq 1$,
  \begin{eqnarray}
\nonumber    \ket{Z_N} &=& e^{-i\frac{J}{2}t}\big(\ket{0}\ket{Z_{N-1}}+2^{\frac{2N-3}{2}}\ket{1}\ket{0}^{\bigotimes N}\big)\\
 &+& e^{i\frac{J}{2}t}\big(-\ket{0}\ket{Z_{N-1}}+2^{\frac{2N-3}{2}}\ket{1}\ket{0}^{\bigotimes N}\big),
\end{eqnarray}
    and
    \begin{eqnarray}
    \bra{Z_N}\ket{Z_N}=2^{2N-1}
    \end{eqnarray}
\textbf{Remark.} Although the method presented here is for the Dicke state with a single excitation, the suitable initial and 
auxiliary qubits (entangled state) can lead to the \(N\)-qubit Dicke state with other excitations  after the evolution. For example, starting with $\ket{\Phi}_{\overline{W}} =\frac{1}{\sqrt{3}}(\ket{011}+\ket{110}+\ket{101})$ and \(N\) auxiliary qubits, we can generate \((3+N)\)-qubit Dicke state with two excitations via Hamiltonian dynamics.

\subsection{Entanglement circulation with imperfect operations}
\label{subsec:disorder}

\begin{figure*}
    \centering
	\includegraphics[width=0.6\textwidth]{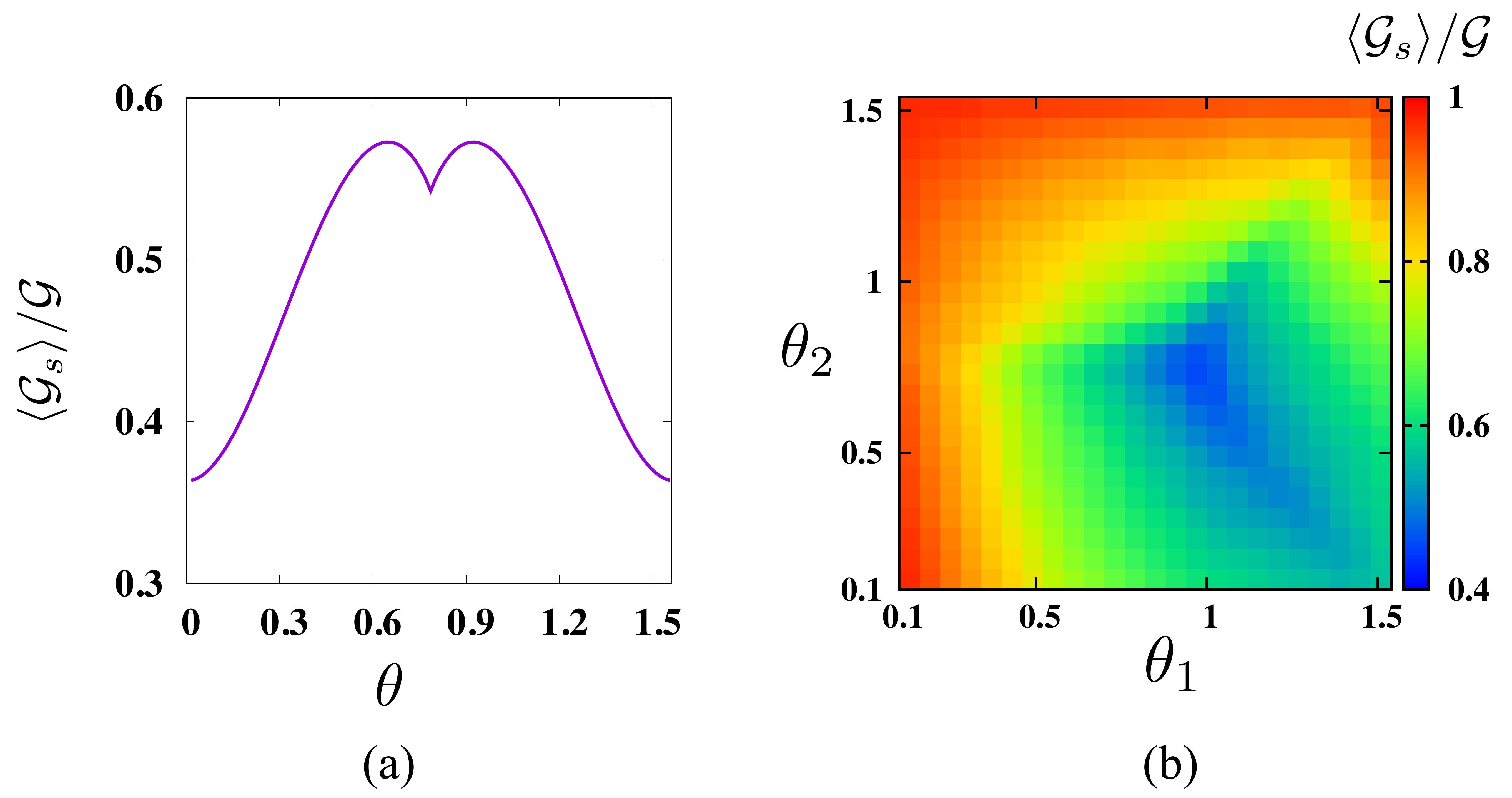}
	\caption{(Color online.) \textbf{Suppression of GGM due to disorder.} (a) Trends of $\langle\mathcal{G}_s\rangle/\mathcal{G}_i$ (vertical axis) as a function of $\theta$ (horizontal axis) when two copies of the gGHZ states, \(\ket{\Phi}_{gGHZ}\) are merged. The evolution is according to the Hamiltonian with \(\langle J \rangle = 0.5 \) and \(\sigma_J = 0.1 \) (b) Same quantities are plotted for \(\ket{\Phi}_{gW}\).  In this case, \(\langle J \rangle = 0.5 \) and \(\sigma_J = 0.3\). In both the cases, we observe that the ratio behaves nonmonotonically with state parameters although they also show differences (see text for details).   Both the axes are dimensionless. } 
	\label{fig:gghz_gw_satur}
\end{figure*}

The  protocol described in the previous subsection works as long as the value of the spin-exchange interaction can be changed instantaneously from zero to a constant value. However, in real situations, there may be fluctuations in the non-zero value of $J$ at $t>0$. Also, a perfect preparation of the  unit states is assumed in the above protocol, which may be difficult to achieve, thereby hindering the efficiency of this protocol. 

We now investigate the effects of such \emph{imperfections} on the performance of this protocol. More specifically, we focus on the time-dependence of GGM of the state $\ket{\Psi(t)}$ in the presence of quenched disorder in $J$, where the time scale of the change of a particular realization is much larger than the evolution time scale of the whole system. Such a disordered system is accessible also in current experimental setups using substrates like cold atoms and trapped ions \cite{Aspect08, deMartini2008, Lewenstein2007, Aspect2009}, which are ideal to create the unit entangled states as well as the two-qubit unitary operators. Moreover, our studies reveal that imperfections can lead to certain advantages in the entanglement properties of the final states.

A quenched disorder in the spin-exchange interaction strength $J$ implies that the time taken by the disordered parameter $J$ to achieve equilibrium is much larger compared to the observation time for the evolution of the system. Therefore, one may consider the value of the disordered parameter to be effectively fixed during the dynamics of the system, thereby making it possible to carry out an averaging of the quantity of interest, $\mathcal{Q}$, over the distribution of different values of the disordered parameters. For randomly chosen spin-exchange interactions $J$ from a probability distribution $P(J)$ with mean $\langle J\rangle$ and standard deviation $\sigma_J$, the quench-averaged $\mathcal{Q}$, denoted by \(\langle \mathcal{Q}\rangle\),  at every time instant is given by
\begin{equation}
\langle \mathcal{Q}\rangle = \int\mathcal{Q}(J)P(J)dJ, \label{eq:dis}
\end{equation}
where $\sigma_J=0$ corresponds to the ordered case discussed in Sec.~\ref{subsec:orderedmodel}. We choose the values of $J$ from a Gaussian distribution with mean \(\langle J\rangle\) and standard deviation \(\sigma_J\).
The recursion relation for arbitrary resource states guarantees that the resulting state for a given realization can also be obtained and hence   we have the functional form  of the integrand in Eq.  (\ref{eq:dis}) for GGM. To investigate the patterns of the quench-averaged GGM for the disordered case, we only numerically compute the integration over the Gaussian distribution.

In contrast with the ordered case discussed in Sec.~\ref{subsec:orderedmodel}, the averaged GGM  after quenching of the six-qubit state $\ket{\Psi(t)}$, originating from either a pair of GHZ-, or a pair of W-, or a GHZ- and a W-class state, is found to oscillate at first, and then saturate to a value $\mathcal{G}_s$ at a critical time $t_c$ (as shown in Fig.~\ref{fig:dis_t_sigma}). This feature is interesting since it exhibits a clear advantage of evolving a system via a disordered Hamiltonian instead of a ordered one. It may also turn out to be important in situations where a quantum protocol requires the GGM of a state to be almost constant over a long period of time. 

The saturation value $\langle\mathcal{G}_s\rangle$ depends on the GGMs of the initial state(s), although no proposition similar to the Proposition III can be put forward to provide a bound on $\langle \mathcal{G}\rangle$. To demonstrate the dependence of $\langle\mathcal{G}_s\rangle$ over the GGM of the resource states, $\mathcal{G}$, we consider two identical copies of a generalized GHZ (gGHZ) state given by $|\Phi\rangle_{gGHZ} = \cos{\theta}|000\rangle + \sin{\theta}|111\rangle$. First of all, we notice that like the initial resource states, \(\langle \mathcal{G}_s\rangle\) increases with the increase of \(\theta\). Towards connecting the saturated value with the initial GGM, we study the trends in the ratio,  $\frac{\langle\mathcal{G}_s\rangle}{\mathcal{G}}$ as $\theta$ is varied. In particular, we observe that the ratio between the saturated value of the quench-averaged GGM and the initial GGM 
increases with an increasing $\theta$ although the increase is not monotonic with \(\theta\) (see Fig.~\ref{fig:gghz_gw_satur}(a)). 
We also test this feature by using generalized W  (gW) states of the form  $|\Phi\rangle_{gW} = \cos{\theta_{1}}|001\rangle+\cos{\theta_{2}}\sin{\theta_{1}}|010\rangle+\sin{\theta_{2}}\sin{\theta_{1}}|100\rangle$ as depicted in 
Fig.~\ref{fig:gghz_gw_satur}(b). 
 The figure shows a clear distinction between the gGHZ and the gW states -- in case of the gGHZ states as inputs, the suppressed averaged value of GGM  for the output state due to disorder decreases  with the increase of the GGM in the inputs while for the gW states as initial, the overall opposite behavior emerges. 

Furthermore,  we also investigate how $t_c$ varies with the distribution of $J$, find it to be decreasing with increasing $\sigma_J$, and eventually saturate at a constant value (see Fig.~\ref{fig:dis_t_sigma}(b)). By employing the \(\chi\)-square curve fitting, we realize that the functional form of \(t_c\) with \(\sigma_J\)  to be \( b + c \exp[-d (\sigma_J -0.01)]\) with \(b=33.2\), \(c=226.2\) and \(d=52.2\) having maximum \(10\%\) errors in parameters.  Note, however,   that for a fixed initial resource states, $\langle\mathcal{G}_s\rangle$ is found to be invariant under a change in the value of $\sigma_J$.

\noindent\textbf{Remark.} Note that we have performed the analysis for disordered operations assuming disorder to be present in one of the evolution operator. Typically in a network, disorder appears in multiple parameters of the Hamiltonian, denoted by $\{x_1,x_2,\cdots,x_n\}$. In such cases, the quenched averaged $\mathcal{Q}$ can be obtained by performing average over different realizations of all these parameters.

\section{Conclusion}
\label{conclusion}

Classical networks are rigorously present to establish communication among different parts of the world, and, on a moderately smaller scale, among multiprocessor devices. However, in this second quantum revolution, the significant advantages of using genuine multiparty entangled states, and in tandem, multiparty quantum networks for performing various quantum information processing tasks are established. Therefore, characterization and implementation of quantum networks play a crucial  role for achieving a communication system with or without security  for the future world. 

In this paper,  we  presented a deterministic protocol, referred to as \emph{entanglement circulation procedure} (ECP), for creating genuine multiparty entangled states, and distributing them in the form of a quantum network. Given a fixed value of entanglement and limited amount of resources, we showed that our  method can generate genuine multiparty entangled states with the application of optimal unitary operators, which is confirmed via computing generalized geometric measure (GGM) of the generated state. Specifically, we proved a bound on the GGM of the resulting state in terms of the GGMs of the initial resource states constituting the network. We also showed that the unitary operators  which can generate maximum GGM in the output state is not unique. Starting from the arbitrary three-qubit initial state, we provided a recursion relation for the output state produced after arbitrary number of steps in this process, thereby spreading genuinely multipartite entangled states in networks. We found that apart from implementing logical gates, these states can also be created  by using interacting quantum spin Hamiltonians. Although we rigorously worked out all the results for a linear geometry of the network, we showed that the ECP  remains equally powerful for other geometries, eg.  triangle-shaped networks. Going beyond the traditional notion of noisy resources, we  considered the scenario where unitary operations are not exact, which can be caused via a disordered spin Hamiltonian.  Counter-intuitively, we observed that although disordered spin Hamiltonian can produce a lesser amount of genuine multipartite entanglement on average compared to the ordered model, in contrast with the latter, the quench-averaged GGM of the resulting state obtained via the evolution of the disordered system can saturate to a constant value after an initial time period. The saturation values of GGM depend on the initial resource states while the strength of the disorder is governed the saturation time. 

The protocol presented in this paper shows an avenue to create genuine multipartite entangled quantum networks. In near future, it will be interesting to find whether all the multipartite resources required for quantum information processing tasks can be generated via this method even in presence of all kinds of noisy environments as well as imperfect operations.

\begin{figure}
	\includegraphics[scale=0.29]{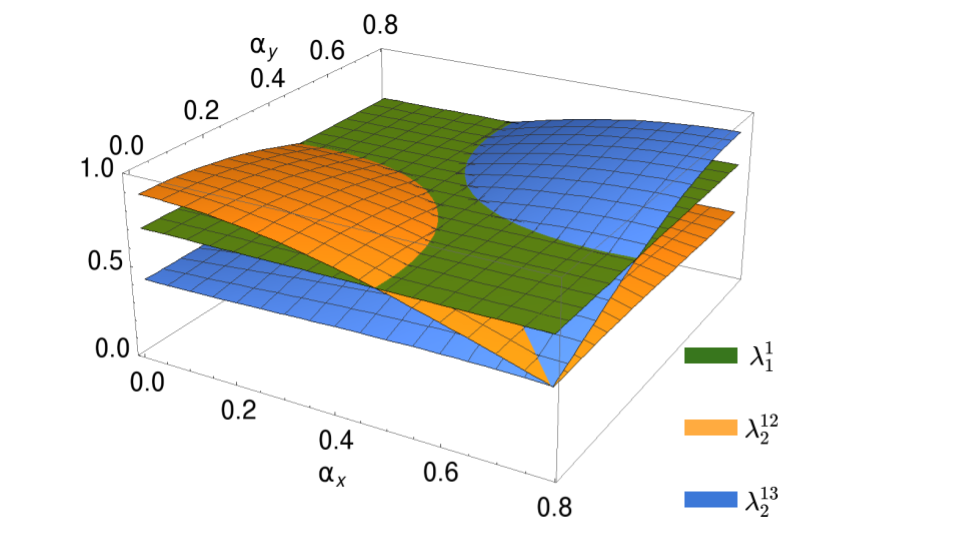}
	\caption{(Color online.) Variations of eigenavlues, $\lambda_1^1$, $\lambda_2^{12}$, and $\lambda_2^{13}$ (\(z\)-axis)  of the output state as functions of $\alpha_x$ (\(x\)-axis) and $\alpha_y$ (\(y\)-axis), with $\alpha_z=0.4$, $\gamma_1=0.7$, and  $\gamma_2=0.6$. See Proposition I for details.
	All the axes are dimensionless. }
	\label{fig:eigenvalues}
\end{figure}

\begin{figure}
	\includegraphics[width=0.8\linewidth]{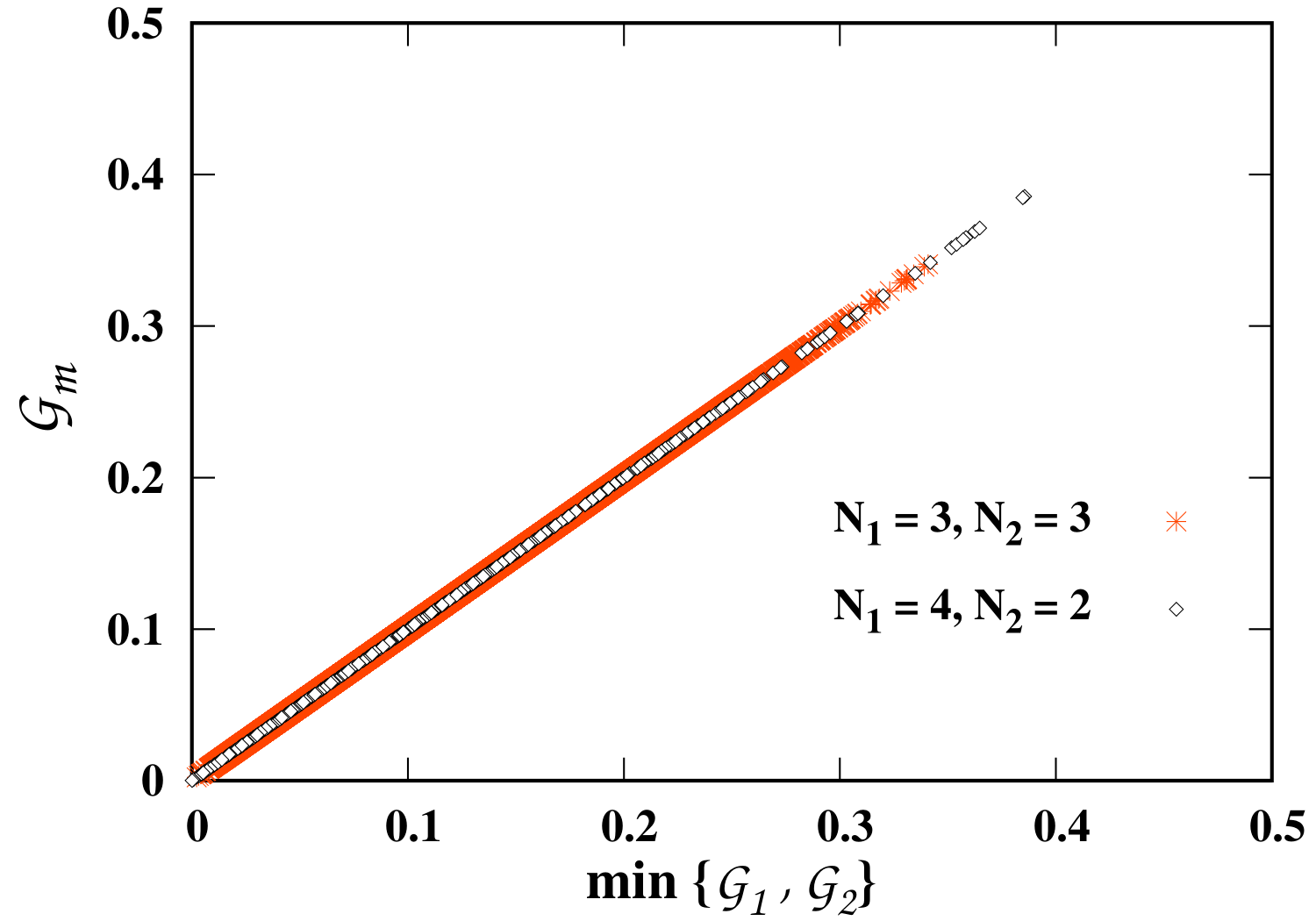}
	\caption{ (Color online.) Scattered plot of maximum output GGM, $\mathcal{G}_m$ (vertical axis) against the initial  GGM of two unit states, $\min \{\mathcal{G}_1, \mathcal{G}_2\}$ (horizontal axis). The output GGM is obtained after optimizing the two-qubit unitary operators, \(\mathcal{U}_2\) when the initial states are Haar uniformly generated (total number of states generated is \(5\times 10^4\)).  The size of the unit cells are either \(N_1 =3, N_2 =3\) (red crosses) or \((N_1 =4, N_2 =2)\) (black diamonds). Both the axes are dimensionless. }
	\label{fig:proof_numerical}
\end{figure}

\acknowledgements

PH, RB, SG, and ASD acknowledge the support from the Interdisciplinary Cyber Physical Systems (ICPS) program of the Department of Science and Technology (DST), India, Grant No.: DST/ICPS/QuST/Theme- 1/2019/23. AKP acknowledges the Seed Grant from IIT Palakkad. We  acknowledge the use of \href{https://github.com/titaschanda/QIClib}{QIClib} -- a modern C++ library for general purpose quantum information processing and quantum computing (\url{https://titaschanda.github.io/QIClib}), and the cluster computing facility at the Harish-Chandra Research Institute. 

\appendix

\section{Generalized geometric measure}
\label{sec:ggmdef}

A $N$-qubit pure state is said to be $k$-\emph{separable} if the multi-qubit state can be written as product of pure states corresponding to $k$-partitions ($2\leq k \leq N$). The geometric measure of entanglement, referred to as the $k$-\emph{geometric measure} ($k$-GM) of entanglement and quantifying multipartite entanglement in a  multi-qubit state $\ket{\Phi}$, is defined as \cite{wei2003, wei2005_a, blasone2008, orus2008, aditi2010}
\begin{eqnarray}
    \mathcal{G}_k=1-\max \left|\langle\Phi_k|\Phi\rangle\right|^2,
\end{eqnarray}
where the maximization is taken over the set of all possible $k$-separable states, $\{\ket{\Phi}_k\}$. For $k=N$, the original definition of the geometric measure of entanglement can be obtained when the maximization is performed over the set of fully separable states \cite{wei2003}. 

On the other hand, $k=2$ corresponds to the \emph{generalized geometric measure} (GGM) of entanglement, which quantifies the maximum distance of the quantum state from the  set of all possible non-genuinely multipartite entangled states. In this paper, we focus on the GGM as the multiparty entanglement quantifier of a multi-qubit state, and denote it with $\mathcal{G}$. It can be shown that in case of GGM, the maximization over the set  $\{\ket{\Phi_k\}}$ reduces to  the maximization over the Schmidt coefficients of all possible bipartitions of $\ket{\Phi}$ \cite{aditi2010}. Mathematically, 
\begin{eqnarray}
    \mathcal{G}=1-\underset{\mathcal{S}_{A:B}}{\max}\{\eta^2\},
\end{eqnarray}
where $\mathcal{S}_{A:B}$ is the full set of all arbitrary bipartitions $A:B$ of the $N$-qubit system, such that $A\cap B=\emptyset$ and $A\cup B=\{1,2,3,\cdots,N\}$, and $\eta$ is the  maximum Schmidt coefficient corresponding to this bipartition.

The above simplification makes the GGM one of the computable multiparty entanglement measure for pure states with arbitrary number of parties in arbitrary dimensions (cf. \cite{GGMmixed1, GGMmixed2}. However, for an $N$-qubit state, one needs to calculate a total of 
$2^{N-1} - 1$  number of reduced density matrices, which increases exponentially with $N$, thereby computing the value of $\mathcal{G}$ difficult for large $N$. Note that the computational challenge can be reduced by restricting to only single- and two-qubit reduced density matrices corresponding to $\ket{\Phi}$ for the computation of $\mathcal{G}$. Note also that while $\mathcal{G}$ remains a multiparty entanglement measure even under this restriction, it may not detect the \emph{genuine} multipartite entanglement in $\ket{\Phi}$. Interestingly, numerical evidences indicate that the value of $\mathcal{G}$ computed in this fashion coincides with the actual value of GGM in several physical systems.

\section{ Extension of Proposition I}
\label{sec:ggm_proposition_numerical}

In situations where the GGM of $\ket{\Psi}_N$ (Eq.~(\ref{eq:effective_four_party state})) is not obtained from the eigenvalues of a single-party density matrix, one needs to investigate all  of the eigenvalues $\{\lambda_1^i,\lambda_2^{jk}\}$, $i=1,2,3,4$, $j=1, k=2,3,4$, (see Sec.~\ref{sec:general_protocol}) obtained from the four single-party and three two-party reduced density matrices. A general analytical treatment is difficult, since different eigenvalues contribute to the computation of GGM for different ranges of the state as well as the unitary parameters. However,
using the proof Proposition I, it is clear that 
\(\mathcal{G} = 1 - \max \{\lambda_1^1, \lambda_2^{12}, \lambda_2^{13}, \lambda_2^{23}\} \). 
Numerical search over appropriate ranges of the relevant parameters indicates that only three among these four eigenvalues, given by 
\begin{widetext}
\begin{eqnarray}
\label{eq:single_party_3}
\lambda_1^1 &=& \gamma_{1},\nonumber\\
\label{eq:two_party_23}
\lambda_2^{13} &=& \frac{1}{4}(1 + (2\gamma_{1} - 1)(2\gamma_{2} - 1)\cos 2\alpha_{x}\cos 2\alpha_{y} + \sin 2\alpha_{x}\sin 2\alpha_{y} + 2\sqrt{F}),\nonumber \\
\label{eq:two_party_13}
\lambda_2^{12} &=& \frac{1}{4}(1 + (2\gamma_{1} - 1)(2\gamma_{2} - 1)\sin 2\alpha_{x}\sin 2\alpha_{y} + \cos 2\alpha_{x} \cos 2\alpha_{y}  + 2\sqrt{G}),
\end{eqnarray}
where
\begin{eqnarray}
F &=& (\gamma_{1} + \gamma_{2} - 1)^{2} \cos^{4}(\alpha_{x} - \alpha_{y}) - 2\cos^{2}(\alpha_{x}-\alpha_{y}) \sin^{2}(\alpha_{x}+\alpha_{y})f + (\gamma_{1} - \gamma_{2})^{2} \sin^{4}(\alpha_{x} + \alpha_{y}),\nonumber \\
G &=& (\gamma_{1} + \gamma_{2} - 1)^{2} \cos^{4}(\alpha_{x} - \alpha_{y}) + 2\cos^{2}(\alpha_{x}-\alpha_{y}) \cos^{2}(\alpha_{x}+\alpha_{y})g + (\gamma_{1} - \gamma_{2})^{2} \cos^{4}(\alpha_{x} + \alpha_{y}), 
\end{eqnarray} 
with
\begin{eqnarray}
f &=& (2\gamma_{1}\gamma_{2} - \gamma_{1} - \gamma_{2})(1+2\gamma_{1}\gamma_{2} - \gamma_{1} - \gamma_{2}) + 4\gamma_{1}\gamma_{2}(1-\gamma_{1})(1-\gamma_{2})\cos 4\alpha_{z}, \nonumber \\ 
g &=& (\gamma_{1} + \gamma_{2}-2\gamma_{1}\gamma_{2})(1+2\gamma_{1}\gamma_{2} - \gamma_{1} - \gamma_{2}) + 4\gamma_{1}\gamma_{2}(1-\gamma_{1})(1-\gamma_{2})\cos 4\alpha_{z},      
\end{eqnarray}
\end{widetext}
contribute in the computation of GGM, subject to the conditions $\gamma_{1} \geq \delta_{1}$, and $\gamma_1 \geq \gamma_{2} \geq  \delta_{2}$ (see Sec.~\ref{sec:general_protocol}). While the eigenvalue in Eq.~(\ref{eq:single_party_3}) emerges from the single-party density matrix corresponding to one of the $2$-dimensional subsystems of $\ket{\Psi}_N$ (Eq.~(\ref{eq:effective_four_party state})), the other two (Eq.~(\ref{eq:two_party_23})) hail from the two-party reduced density matrices. We consider the possible ranges of $\gamma_1$ and $\gamma_2$ within the normalization condition of the pair of states $\ket{\Phi}_{N_1}$ and $\ket{\Phi}_{N_2}$. For each pair of states $\ket{\Phi}_{N_1}$ and $\ket{\Phi}_{N_2}$, we numerically search for a set of values of $\{\alpha_x,\alpha_y,\alpha_z\}$, such that the GGM of  $\ket{\Psi}_N=U_d\left[\ket{\Phi}_{N_1}\otimes\ket{\Phi}_{N_2}\right]$ is obtained from a single-qubit reduced density matrix. We find the search to be successful for all pairs of states in the set, and for each pair, multiple instances of $U_d$ are found. As a demonstration, consider  Fig.~\ref{fig:eigenvalues}, where we plot the variations of the eigenvalues in Eq.~(\ref{eq:single_party_3}) as functions of $\alpha_x$ and $\alpha_y$ for typical fixed values of $\alpha_z$, $\gamma_1$, and $\gamma_2$. The figure clearly indicates that $\lambda_1^1$ is the minimum among the three eigenvalues for a set of values of $\alpha_{j}$, $j=x,y,z$, implying that the GGM of $\ket{\Psi}_N$ is $1-\lambda_1^1$, thereby validating Proposition III.  

The above numerical search assumes that the GGMs of the unit states, $\ket{\Phi}_{N_1}$ and $\ket{\Phi}_{N_2}$, are obtained from the single-qubit density matrices (see Eqs.~(\ref{eq:unit_state_ggm_1})-(\ref{eq:unit_state_ggm_2}) in Sec.~\ref{sec:general_protocol}). In order to check whether the GGM of the resultant state is given by the minimum of the GGMs of the unit states even when this assumption is relaxed, we Haar-uniformly generate two-, three-, and four-qubit quantum states (a sample of size \(5 \times 10^4\) in each case) to produce six-qubit resultant states, and check the Proposition III for pairs of states from these sets. Our numerical result suggests that Proposition III holds in all of these cases, which is also demonstrated in Fig.~\ref{fig:proof_numerical}. We elaborate more on the implications of this numerical analysis in Sec.~\ref{subsec:optimal_resource_distribution}.     

\section{Proof of Proposition IV}
\label{sec:recursion_derivation}

Here we derive the recursion relation describing the form of the $3m$-qubit state $\ket{\Psi}_{3m}$, constituted of $m$ identical three-qubit pure unit states by applying $m-1$ unitary operators $\{\mathcal{U}_2^j;j=1,2,\dots,m-1\}$ (see \textbf{Proposition IV}). Let us first consider two identical three-qubit states of the form
\begin{eqnarray}
\ket{\Phi} &=& \sum_{i=1}^8a_{i}\ket{b_i},
\label{GHZboom_app}
\end{eqnarray}
where $\{a_{i} \in \mathbb{C}\;\forall i\}$, and $\{\ket{b_i}\}$ is the product basis for three-qubits, constituted of the single-qubit computational basis. The initial state of the six-qubit system is given by $\ket{\Phi}_{(123456)}=\ket{\Phi}_{(123)}\otimes\ket{\psi}_{(456)}$, where in our notations, subscripts and superscripts in brackets respectively for states and unitary operators represent the labels of the qubits (see the first two blocks of Fig.~\ref{fig:schem2}), and we have temporarily dropped the number of qubits from the subscripts of the unit states for brevity. It is convenient to write $\ket{\Phi}$ as    
\begin{widetext}
\begin{eqnarray}
\ket{\Phi}_{(123456)} &=& \ket{A}_{(12)}\ket{0}_{(3)}+\ket{B}_{(12)}\ket{1}_{(3)}
=\ket{0}_{(1)}\ket{E}_{23}+\ket{1}_{(1)}\ket{F}_{(23)},
\label{GHZkaboom_app}
\end{eqnarray}
where
\begin{eqnarray}
\ket{A} &=& a_{1}\ket{00}+a_{2}\ket{01}+a_{4}\ket{10}+a_{7}\ket{11},\;
\ket{B} = a_{3}\ket{00}+a_{5}\ket{01}+a_{6}\ket{10}+a_{8}\ket{11},\nonumber\\
\ket{E} &=& a_{1}\ket{00}+a_{3}\ket{01}+a_{2}\ket{10}+a_{5}\ket{11},\;
\ket{F} = a_{4}\ket{00}+a_{6}\ket{01}+a_{7}\ket{10}+a_{8}\ket{11}.
\end{eqnarray}
\end{widetext}
Noting that applying the unitary operator $U_d$ is enough to investigate entanglement of the resulting state $\ket{\Psi}_{(123456)}$, and applying $U_d^1$ on qubits $3$ and $4$ (see Fig.~\ref{fig:schem2}) where the superscript "$1$" represent the value of the unitary index $j$ in $\mathcal{U}_2$ (see Secs.~\ref{subsec:protocol} and \ref{sec:three_qubit_units}),
\begin{widetext}\small 
\begin{eqnarray}
\ket{\Psi}_{(123456)} &=& \mathcal{U}_{(34)}^1\ket{\Phi}_{(123456)}\nonumber \\
&=&\mathcal{U}_{d(34)}^1 \left[( \ket{A}_{(12)}\ket{0}_{(3)}+\ket{B}_{(12)}\ket{1}_{(3)}) \otimes (\ket{0}_{(4)}\ket{E}_{(56)}+\ket{1}_{(4)}\ket{F}_{(56)})\right]\nonumber \\
&=& \left[\ket{A}_{(12)}U_{d(34)}^1\ket{00}_{(34)}+\ket{B}_{(12)}U_{d(34)}^1\ket{10}_{(34)}\right]\ket{E}_{(56)}
+\left[\ket{B}_{(12)}U_{d(34)}^1\ket{11}_{(34)}+\ket{A}_{(12)}U_{d(34)}^1\ket{01}_{(34)}\right]\ket{F}_{(56)} \nonumber\\
&=& \ket{X}^1_{1234}\ket{E}_{(56)}+\ket{Y}^1_{1234}\ket{F}_{(56)},
\end{eqnarray}\normalsize 
where
\begin{eqnarray}
\ket{X}^1_{1234} &=&   \ket{A}_{(12)}U_{d(34)}^1\ket{00}_{(34)}+\ket{B}_{(12)}U_{d(34)}^1\ket{10}_{(34)},\;
\ket{Y}^1_{1234} = \ket{B}_{(12)}U_{d(34)}^1\ket{11}_{(34)}+\ket{A}_{(12)}U_{d(34)}^1\ket{01}_{(34)},\nonumber \\
\end{eqnarray}
\end{widetext}
and the superscripts to the states $\ket{X}$ and $\ket{Y}$ represent the number of $\mathcal{U}_2$ operators applied so far. Moving a step further and applying $\mathcal{U}_2$ on the qubits $6$ and $7$ in the state $\ket{\Psi}_{(123456)}\ket{\psi}_{(789)}$, one obtains an $9$-qubit state as (see Fig.~\ref{fig:schem2})
\begin{eqnarray}
\ket{\Psi}_{(123456789)} = \ket{X}^2_{(1234567)}\ket{E}_{(89)}+\ket{Y}^2_{(1234567)}\ket{F}_{(89)},\nonumber\\ 
\end{eqnarray}
where
\begin{widetext}
\small 
\begin{eqnarray}
\ket{X}^2_{(1234567)} &=& \left[\ket{X}^1_{(1234)}(a_{1}\ket{0}+a_{2}\ket{1})_{(5)} 
+\ket{Y}^1_{(1234)}(a_{4}\ket{0}+a_{7}\ket{1})_{(5)}\right]U_{d(67)}^2\ket{00}_{(67)}\nonumber\\
&&+\left[\ket{X}^1_{(1234)}(a_{3}\ket{0}+a_{5}\ket{1})_{(5)} + \ket{Y}^1_{(1234)}(a_{6}\ket{0}+a_{8}\ket{1})_{(5)}\right]U_{d(67)}^2\ket{10}_{(67)},\\
\ket{Y}^2_{(1234567)} &=&
\left[\ket{X}^1_{(1234)}(a_{1}\ket{0}+a_{2}\ket{1})_{(5)}+\ket{Y}^1_{(1234)}(a_{4}\ket{0}+a_{7}\ket{1})_{(5)}\right]U_{d(67)}^2\ket{01}_{(67)} \nonumber \\
&&+\left[\ket{X}^1_{(1234)}(a_{3}\ket{0}+a_{5}\ket{1})_{(5)}
+\ket{Y}^1_{(1234)}(a_{6}\ket{0}+a_{8}\ket{1})_{(5)}\right]U_{d(67)}^2\ket{11}_{(67)}.
\end{eqnarray}
\normalsize 
\end{widetext}
This procedure can be continued for an arbitrary number of three-qubit states belonging to the GHZ-class, where after applying $l$ unitary operators $\mathcal{U}_2$, a multiparty state of $3(l+1)$ qubits having the form  
\begin{widetext}
\begin{equation}
\ket{\Psi}_{3(l+1)} = \ket{X}^{l}\ket{E}+\ket{Y}^{l}\ket{F}
\end{equation}
is obtained. Here, 
\begin{eqnarray}
\ket{X}^{l} &=&\left[\ket{X}^{l-1}(a_{1}\ket{0}+a_{2}\ket{1})+\ket{Y}^{l-1}(a_{4}\ket{0}+a_{7}\ket{1})\right]U_d^{l}\ket{00}\nonumber\\ 
&&+\left[\ket{X}^{l-1}(a_{3}\ket{0}+a_{5}\ket{1})+\ket{Y}^{l-1}(a_{6}\ket{0}+a_{8}\ket{1})\right]U_d^{l}\ket{10},\nonumber\\
\ket{Y}^{l} &=&\left[\ket{X}^{l-1}(a_{1}\ket{0}+a_{2}\ket{1})+\ket{Y}^{l-1}(a_{4}\ket{0}+a_{7}\ket{1})\right]U_d^{l}\ket{01}\nonumber\\ 
&&+\left[\ket{X}^{l-1}(a_{3}\ket{0}+a_{5}\ket{1})+\ket{Y}^{l-1}(a_{6}\ket{0}+a_{8}\ket{1})\right]U_d^{l}\ket{11}, 
\end{eqnarray}
where each of the states $\ket{X}^{l}$ and $\ket{Y}^l$ can be derived for an arbitrary value of $l$ starting from 
\begin{eqnarray}
\ket{X}^1 &=& \ket{A}U_d^1\ket{00}+\ket{B}U_d^1\ket{10},\;
\ket{Y}^1 = \ket{B}U_d^1\ket{11}+\ket{A}U_{d}^1\ket{01}.\nonumber \\
\end{eqnarray}
Clearly, for $\ket{\Psi}_{3m}$, $l=m-1$.
\end{widetext}

\bibliography{ref}
\end{document}